\documentclass[twocolumn,english,aps,superscriptaddress,tightenlines,notitlepage,amssymb]{revtex4-2}
\usepackage{lmodern}
\usepackage[T1]{fontenc}
\usepackage[latin9]{inputenc}
\usepackage{color}
\usepackage{babel}
\usepackage{amsmath}
\usepackage{amssymb}
\usepackage{cancel}
\usepackage{graphicx}
\usepackage{esint}
\usepackage[unicode=true,pdfusetitle,
 bookmarks=true,bookmarksnumbered=true,bookmarksopen=false,
 breaklinks=false,pdfborder={0 0 1},backref=false,colorlinks=true]
 {hyperref}
\hypersetup{
 linkcolor=blue,urlcolor=blue,citecolor=blue,pdfstartview={FitH},hyperfootnotes=false,bookmarks=False}

\makeatletter

\providecommand{\tabularnewline}{\\}

\usepackage{babel}
\date{\today}
\usepackage{xcolor}
\usepackage{graphicx}
\allowdisplaybreaks

\renewcommand{\big}{\bBigg@\@ne}
\renewcommand{\Big}{\bBigg@{1.5}}
\renewcommand{\bigg}{\bBigg@\tw@}
\renewcommand{\Bigg}{\bBigg@{2.5}}

\newcommand{\biggg}{\bBigg@\thr@@}
\newcommand{\Biggg}{\bBigg@{3.5}}

\makeatother

\begin{document}
\title{Linear optical response from the \emph{odd} parity Bardasis-Schrieffer
mode in locally non-centrosymmetric superconductors}
\author{Changhee Lee}
\affiliation{Department of Physics and Astronomy, Seoul National University, Seoul
08826, Korea}
\author{Suk Bum Chung}
\email{sbchung0@uos.ac.kr}

\affiliation{Department of Physics and Natural Science Research Institute, University
of Seoul, Seoul 02504, Korea}
\affiliation{School of Physics, Korea Institute for Advanced Study, Seoul 02455,
Republic of Korea}
\begin{abstract}
On the recent report of a magnetic field induced first order transition
between an even-parity superconductivity and an odd-parity superconductivity
in $\mathrm{CeRh_{2}As_{2}}$ {[}Khim \emph{et al}. \href{https://www.science.org/doi/10.1126/science.abe7518}{Science, 373, 1012 (2021)}{]},
the microscopic physics is still under investigation. However, if,
in the vicinity of this transition, the coupling strengths of the
even and odd pairing channels are comparable, a particle-particle
excitonic collective mode referred to as the Bardasis-Schrieffer (BS)
mode should generically exist below the pair-breaking continuum. This
BS mode can couple to the light and thus affect the optical response
of the superconductor, as it arises from a pairing channel with the
parity opposite to that of the ground state pairs. Here, by using
a generic bilayer model Hamiltonian for the electronic degree of freedom,
which is globally centrosymmetric despite each layer being locally
non-centrosymmetric, we study the change of the excitation gap of
the BS mode with respect to the out-of-plane magnetic fields and demonstrate
that its coupling to the light is possible even in the linear response
regime. The linear coupling is attributed to the presence of multiple
electronic bands, which is a generic feature of a bilayer system.
Our result shows the microwave absorption as the signature of the
BS mode, and hence a smoking gun signature of the parity-switching
at the transition between two superconducting phases.
\end{abstract}
\maketitle

\section{Introduction}

Discovering superconductors with odd-parity Cooper pairing has been
a long-standing challenge in condensed matter physics, as they are
rare in inversion symmetric solid state systems. To name a few, ${\rm UPt_{3}}$
\citep{Joynt2002}, ${\rm UNi_{2}Al_{3}}$ \citep{Ishida2002} and
${\rm Sr_{2}RuO_{4}}$ \citep{Mackenzie2017} are the most notable
candidates which have been suspected to host odd-parity Cooper pairings
for a long time, though the case of a much studied candidate material
${\rm Sr_{2}RuO_{4}}$ has grown more controversial in recent years
\citep{Pustogow2019,Ishida2020,Petsch2020,Suh2020,Kaeser2022}.

Faced with this rarity of odd-parity superconducting materials, many
research have endeavored to find realistic conditions favoring the
odd parity superconductivity. For instance, the systems possessing
a structural instability toward an inversion-symmetry-broken phase
such as the pyrochlore oxide ${\rm Cd_{2}Re_{2}O_{7}}$ drew attention
for a potential to host an odd-parity superconducting phase \citep{Kozii2015,Schumann2020,Wang2016,Wang2017}.

Another mechanism for odd-parity superconductivity is suggested by
a recent experiment on $\mathrm{CeRh_{2}As_{2}}$ \citep{Khim2021,Hafner2022}.
There a transition is observed when the external magnetic fields are
applied along the $c$-axis within the superconducting phase of $\mathrm{CeRh_{2}As_{2}}$
\citep{Khim2021}. According to the preceding theoretical studies
\citep{Maruyama2012,Sigrist2014}, the transition referred to as the
even-to-odd transition seems to occur between two superconducting
phases of opposite parities under an inversion. The Pauli paramagnetic
pair-breaking effect \citep{Clogston1962,Sarma1963} is a known mechanism
for destroying the even-parity superconducting (eSC) phase. By contrast,
an odd-parity superconducting (oSC) state can withstand the magnetic
fields through an equal-(pseudo)spin pairing \citep{Maeno2012,Khim2021,Cavanagh2022,Mockli2021}.
It is noted that the combination of\emph{ $P4/nmm$} nonsymmorphic
crystal structure and the heavy-fermion characteristic supports strong
intralayer Rashba-type spin-orbit couplings that are known to favor
equal-(pseudo)spin pairings \citep{Cavanagh2022}.

An intriguing implication of the even-to-odd transition in $\mathrm{CeRh_{2}As_{2}}$
is that the coupling strengths of the attractive interactions for
both pairing channels may be comparable. The potential transition
temperature $T_{c,o}$ of the oSC phase at the zero-field, which is
preempted by the onset of the eSC phase in reality, is estimated to
be close to the transition temperature $T_{c,e}$ of the eSC phase
\citep{Khim2021}. Moreover, phenomenological studies have reproduced
the overall superconducting phase diagram in $\mathrm{CeRh_{2}As_{2}}$
with comparable coupling strengths for both pairing channels \citep{Khim2021,Yoshida2012}.

Even if the most of the theories set forth so far support that the
high-field superconducting phase of ${\rm CeRh_{2}As_{2}}$ is odd
in parity, counter-arguments have also been raised. For instance,
a theoretical study proposed that the observed magnetic field induced
phase transition arises not from the parity switching of the superconducting
gap but from the spin-flopping in the coexistent antiferromagnetic
order parameter \citep{Machida2022}. Therefore, further experimental
signatures need to be sought for the first-order transition that switches
the parity of the superconducting gap. Of the many ways to find an
indisputable evidence for the symmetry of the superconducting phase,
one is to investigate the collective modes in the superconducting
phase. Historically, the detection of a number of the nearly-degenerate
collective modes in the superfluid $B$-phase of ${\rm ^{3}He}$ proved
to be the decisive evidence in favor of the spin-triplet pairing \citep{Sauls2022}.

In this regards, we note that, if the even-to-odd transition is really
a parity-switching transition, $T_{c,o}\approx T_{c,e}$, which implies
the close competition between two pairing channels of opposite parities,
provides a favorable condition for a collective mode, known as the
Bardasis-Schrieffer (BS) mode \citep{Bardasis1961}, to appear far
from the pair-breaking continuum. The BS mode is an exciton-like collective
mode in superconductors due to a sub-leading pairing channel and indicates
an instability towards another superconducting phase breaking some
symmetries of the superconducting ground state. As a precursor of
the instability of the superconducting ground state, the gap of the
BS mode becomes smaller as the sub-leading channel gets stronger.
However, such closely competing pairing channels have been rarely
been found in superconductors, with one of a few exceptions being
the iron-based superconductors, where the close competition between
the $s$-wave and $d$-wave pairing channels have been confirmed by
the Raman detection of the BS mode \citep{Bohm2014,He2020}.

Besides the possible existence of the BS mode, it is worth noting
that the collective mode can possess a non-zero optical coupling when
the parity of the sub-leading pairing channel under inversion is the
opposite of that of the superconducting ground state. This feature
makes the detection of the collective mode possible through the optical
response in the \emph{linear} response regime, which can be thought
of as a compelling proof for the existence of a strong odd-parity
pairing channel. This is in a sharp contrast to the Fe-based superconductors
where the electronic Raman spectroscopy is used to detect the BS mode
from the $d$-wave channel as this pairing channel and the $s$-wave
ground state pairing share the same parity \citep{Bohm2014,He2020}.
Thus, in the case of ${\rm CeRh_{2}As_{2}}$, the detection of the
BS mode would be a smoking gun evidence for the occurrence of the
parity-switching at the observed transition between the two superconducting
phases.

In this work, we conduct a qualitative study on the BS modes in the
clean limit superconducting phase of a locally non-centrosymmetric
system such as $\mathrm{CeRh_{2}As_{2}}$, which arise from the odd-parity
and even-parity pairing channel in the eSC state and oSC state, respectively.
First, we demonstrate the even-to-odd parity transition by the Pauli
paramagnetic effect at the zero-temperature at the level of a mean-field
description. We then briefly introduce the generalized random phase
approximation (GRPA) \citep{Anderson1958,Rickayzen1959} which provides
the basis of the analysis in this work. Also, it is shown that the
BS modes from the subdominant pairing channels can be linearly coupled
to the light. This is ascribed to the origin of the BS mode whose
parity is opposite to the ground state Cooper pairing. Using the GRPA,
we investigate the softening of the BS modes under the external magnetic
fields along $c$-axis and the linear optical response from the BS
modes.

\section{First order transition by Pauli paramagnetic pair-breaking}

To discuss the BS mode under the external magnetic field, the critical
magnetic fields for the even-to-odd transition in the superconducting
phase should be found first. Thus, we start our presentation by demonstrating
the even-to-odd transition in the superconducting phase in a locally
non-centrosymmetric layered structure by using a mean-field description
at the zero-temperature. For results valid in a more wide range of
temperature and magnetic fields, we refer to Refs. \citep{Maruyama2012,Sigrist2014,Mockli2021,Mockli2018}.

\begin{figure}
\includegraphics[width=0.9\columnwidth]{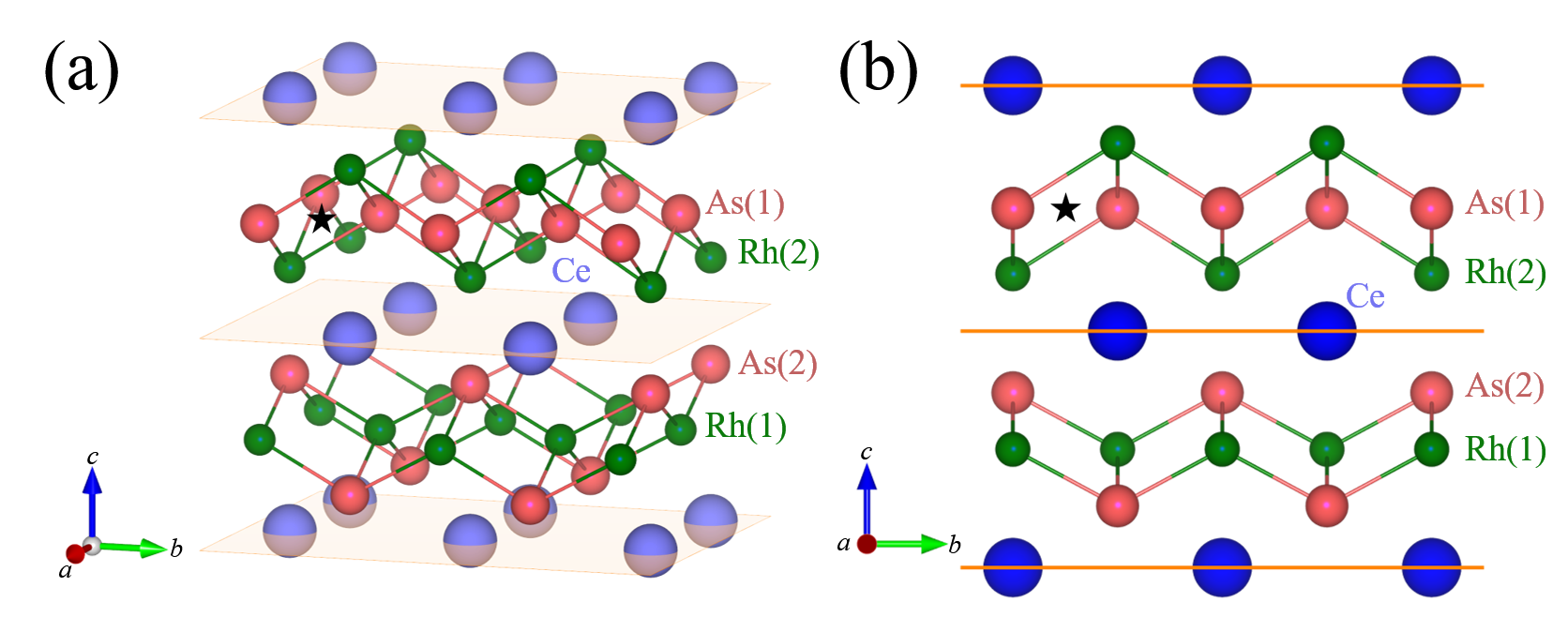}

\caption{\label{fig:crystal}Crystal structure of ${\rm CeRh_{2}As_{2}}$.
(a) Bird's eye view of the structure. (b) A view from the (100) direction.
An inversion center is marked by a black star.}
\end{figure}

Let us begin with a model Hamiltonian for the\emph{ }normal phase
of the representative locally non-centrosymmetric system, ${\rm CeRh_{2}As_{2}}$,
subject to the point group $D_{4h}$ which is given by \citep{Sigrist2014,Mockli2021,Khim2021,Cavanagh2022}:
\begin{align}
H_{0}(\boldsymbol{k})= & \sum_{i=0}^{2}\varepsilon_{i0}(\boldsymbol{k})\sigma_{i}+\sum_{i=1}^{3}\varepsilon_{3i}(\boldsymbol{k})\sigma_{3}s_{i}-\mu\label{eq:CeRhAs-normal}
\end{align}
 with 
\begin{align}
\varepsilon_{00}(\boldsymbol{k}) & =2t(2-\cos k_{x}-\cos k_{y}),\\
\varepsilon_{10}(\boldsymbol{k}) & =t_{c,1}\cos\frac{k_{z}}{2}\cos\frac{k_{x}}{2}\cos\frac{k_{y}}{2},\\
\varepsilon_{20}(\boldsymbol{k}) & =t_{c,2}\sin\frac{k_{z}}{2}\cos\frac{k_{x}}{2}\cos\frac{k_{y}}{2},\\
\varepsilon_{31}(\boldsymbol{k}) & =-\alpha_{\text{R}}\sin k_{y},\\
\varepsilon_{32}(\boldsymbol{k}) & =\alpha_{\text{R}}\sin k_{x},\\
\varepsilon_{33}(\boldsymbol{k}) & =\lambda_{\text{I}}\sin k_{z}\sin k_{x}\sin k_{y}(\cos k_{x}-\cos k_{y}),
\end{align}
where $\sigma_{i}$ and $s_{i}$ are the Pauli matrices for the orbital
and spin degrees of freedom, respectively. Here, two orbital degrees
of freedom are introduced to take account of the locally non-centrosymmetric
feature of the system. The reason is easily understood by looking
into the crystal structure of ${\rm CeRh_{2}As_{2}}$ drawn in Figure
\ref{fig:crystal}. In Fig. \ref{fig:crystal}(a), the crystal structure
is depicted with three \{001\} lattice planes composed of ${\rm Ce}$
atoms. The locally broken inversion symmetry around ${\rm Ce}$ atoms
is easily noted in Fig. \ref{fig:crystal}(b) where the crystal structure
is viewed from the (100) direction. The black stars in Fig. \ref{fig:crystal}
correspond to a center for the global inversion symmetry, under which
no individual atom is left invariant.This global inversion is represented
by ${\cal P}=\sigma_{1}s_{0}$ in the basis of the model Hamiltonian
$H_{0}(\boldsymbol{k})$ in Eq. \eqref{eq:CeRhAs-normal}.

$t_{c,1}$ and $t_{c,2}$ are the hoppings between the nearest-neighbor
${\rm Ce}$ layers depicted in Fig. \ref{fig:crystal}. These hoppings
endow the three-dimensional characteristics to the electronic structure.
$\alpha_{\text{R}}$ and $\lambda_{\text{I}}$ denote the intra-layer
Rashba- and inter-layer Ising-type spin-orbit couplings, respectively.
Note that the sign of the Rashba spin-orbit coupling alternate layer
by layer, which reflects the locally non-centrosymmetric structure
of the system shown in Fig. 1(b).

Throughout this work, we ignore $\lambda_{\text{I}}$ since this spin-orbit
coupling corresponds to a spin-dependent inter-layer hopping between
the two \emph{next-}nearest-neighboring layers, and thus it is expected
to be much weaker than the spin-independent inter-layer hoppings $t_{c,1}$,
$t_{c,2}$ between the nearest layers and the intra-layer Rashba spin-orbit
coupling $\alpha_{\text{R}}$. Also, we assume that the Rashba-type
spin-orbit coupling $\alpha_{\text{R}}$ is much larger than the intra-layer
hoppings $t_{c,1}$ and $t_{c,2}$ following Refs. \citep{Cavanagh2022,Khim2021}.
In this limit of large Rashba spin-orbit coupling, the difference
between $t_{c,1}$ and $t_{c,2}$ has no significant effect on the
band structure except for a weak modulation of the Fermi surface along
the $k_{z}$-axis. Thus, $t_{c}\equiv t_{c,1}=t_{c,2}$ is assumed
throughout this work.

The two-fold degenerate eigenenergies of $H_{0}(\boldsymbol{k})$
in Eq. \eqref{eq:CeRhAs-normal} are given by
\begin{align}
\xi_{1}(\boldsymbol{k}) & =\varepsilon_{00}(\boldsymbol{k})-\mu+\sqrt{t(\boldsymbol{k})^{2}+\alpha(\boldsymbol{k})^{2}},\\
\xi_{2}(\boldsymbol{k}) & =\varepsilon_{00}(\boldsymbol{k})-\mu-\sqrt{t(\boldsymbol{k})^{2}+\alpha(\boldsymbol{k})^{2}},
\end{align}
with $t(\boldsymbol{k})=\sqrt{\varepsilon_{10}(\boldsymbol{k})^{2}+\varepsilon_{20}(\boldsymbol{k})^{2}}$
and $\alpha(\boldsymbol{k})=\sqrt{\varepsilon_{31}(\boldsymbol{k})^{2}+\varepsilon_{32}(\boldsymbol{k})^{2}}$.
The eigenvectors of these eigenvalues are
\begin{subequations}
\begin{align}
|\xi_{1},\alpha\rangle=\left(\begin{matrix}\frac{\cos\frac{\chi}{2}}{\sqrt{2}}\\
\frac{e^{i\phi}\sin\frac{\chi}{2}}{\sqrt{2}}\\
\frac{e^{i\zeta}\cos\frac{\chi}{2}}{\sqrt{2}}\\
\frac{z\sin\frac{\chi}{2}}{-\sqrt{2}}
\end{matrix}\right),\; & |\xi_{1},\beta\rangle=\left(\begin{matrix}\frac{\sin\frac{\chi}{2}}{z\sqrt{2}}\\
\frac{e^{-i\zeta}\cos\frac{\chi}{2}}{\sqrt{2}}\\
\frac{e^{-i\phi}\sin\frac{\chi}{2}}{-\sqrt{2}}\\
\frac{\cos\frac{\chi}{2}}{\sqrt{2}}
\end{matrix}\right),\label{eq:CeRhAs_Eigenvectors1}\\
|\xi_{2},\alpha\rangle=\left(\begin{matrix}\frac{\cos\frac{\chi}{2}}{z\sqrt{2}}\\
\frac{e^{-i\zeta}\sin\frac{\chi}{2}}{-\sqrt{2}}\\
\frac{e^{-i\phi}\cos\frac{\chi}{2}}{-\sqrt{2}}\\
\frac{\sin\frac{\chi}{2}}{-\sqrt{2}}
\end{matrix}\right),\; & |\xi_{2},\beta\rangle=\left(\begin{matrix}\frac{\sin\frac{\chi}{2}}{\sqrt{2}}\\
\frac{e^{i\phi}\cos\frac{\chi}{2}}{-\sqrt{2}}\\
\frac{e^{i\zeta}\sin\frac{\chi}{2}}{\sqrt{2}}\\
\frac{z\cos\frac{\chi}{2}}{\sqrt{2}}
\end{matrix}\right),\label{eq:CeRhAs_Eigenvectors2}
\end{align}
\end{subequations}
 with $\exp(i\chi)=\{t(\boldsymbol{k})+i\alpha(\boldsymbol{k})\}/\sqrt{t(\boldsymbol{k})^{2}+\alpha(\boldsymbol{k})^{2}}$,
$\exp(i\zeta)=\{\varepsilon_{10}(\boldsymbol{k})+i\varepsilon_{20}(\boldsymbol{k})\}/t(\boldsymbol{k}),$
$\exp(i\phi)=\{\varepsilon_{31}(\boldsymbol{k})+i\varepsilon_{32}(\boldsymbol{k})\}/\alpha(\boldsymbol{k})$,
and $z=\exp[i(\zeta+\phi)]$.

\begin{figure}
\centering\includegraphics[width=1\columnwidth]{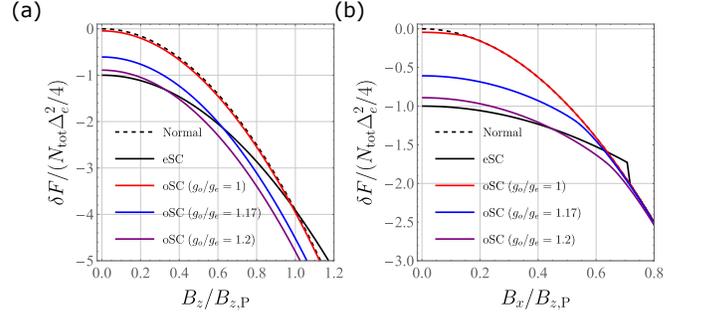}\caption{\label{fig:fig1}Free energy comparison of the normal phase and the
superconducting phases. (a) Under the out-of-plane magnetic fields.
(b) Under the in-plane magnetic fields. For comparison, the horizontal
axes of both figures are normalized by the same $B_{z,\text{P}}$.
The dashed line and the solid black line represent the free energies
of the normal phase and the eSC phase, respectively. They meet at
$B_{z}=B_{z,\text{P}}$. The red, blue, purple lines are the free
energies of oSC states for $g_{o}/g_{e}=1,\;1.17,\;1.2$. When these
lines (oSC) cross the black solid line (eSC), the even-to-odd transition
occurs. The parameters $t=2$, $\mu=0.5$, $t_{c,1}=t_{c,2}=0.1$,
$\alpha_{\text{R}}=0.34$ and $\Delta_{e}=0.004$ are used. The ratio
of $\alpha$ and $t_{c,1}=t_{c,2}$ is adopted from Refs. \citep{Cavanagh2022,Khim2021}.}
\end{figure}

Provided this lattice model, we assume that two pairing channels whose
form factors are represented by $\sigma_{0}$ and $\sigma_{z}$ are
the predominant pairing channels in eSC and oSC states, respectively,
which are used to reproduce the $H-T$ phase diagram of ${\rm CeRh_{2}As_{2}}$
in Refs. \citep{Cavanagh2022,Khim2021,Mockli2018,Mockli2021}. The
gap function $\sigma_{0}$ in the eSC state is uniform while the sign
of the gap function $\sigma_{z}$ alternates layer-by-layer in the
oSC state. Including the Zeeman term $\boldsymbol{B}\cdot\boldsymbol{s}$,
the Bogoliubov-de Gennes (BdG) Hamiltonians for the eSC and oSC states
are given by
\begin{equation}
H_{\text{BdG}}^{(p)}(B,\boldsymbol{k})=\tau_{0}\boldsymbol{B}\cdot\boldsymbol{s}+\tau_{z}H_{0}(\boldsymbol{k})+\Delta_{p}\tau_{x}^{(p)},\label{eq:CeRhAs-BdG}
\end{equation}
with $p=e,o$ and $\tau_{i}^{(p)}=\tau_{i}M_{p}$ for $i=x,y,z$ where
$M_{e}=\sigma_{0}$ and $M_{o}=\sigma_{z}$ are the pairing form factors
, respectively. The magnetic field along (perpendicular) the $z$-axis,
which corresponds to the $c$-axis of ${\rm CeRh_{2}As_{2}}$, is
denoted by $B_{z}$($B_{x}$) and it is referred to as the out-of-plane
(in-plane) magnetic field in this work. Here, the basis field operator
of the BdG Hamiltonian is $\hat{\Psi}_{\boldsymbol{k}}=(\hat{C}_{\boldsymbol{k}},\hat{C}_{-\boldsymbol{k}}^{\dagger T}(is_{y}))^{T}$
\citep{Balian1963,Nambu1960}. The gap amplitude $\Delta_{p}$ presumed
to be real is determined from the gap equation:
\begin{equation}
\Delta_{p}=\frac{g_{p}}{2}\check{\sum_{k}}\text{Tr}[\tau_{x}^{(p)}G_{k}^{(p)}].\label{eq:GapEq}
\end{equation}
with $G_{k}^{(p)}=i\omega-H_{\text{BdG}}^{(p)}(\boldsymbol{k})$.
Here, $\check{\sum}_{k}=(\beta V)^{-1}\sum_{k}$ is the normalized
summation over $k=(i\omega,\boldsymbol{k})$ a pair of Matsubara frequency
and the three-dimensional momentum, where $\beta=1/k_{B}T$ and $V$
are the inverse of the temperature and the volume of the system, respectively.
The coupling constants $g_{e}$ and $g_{p}$ are assumed to be constant
for the simplicity of the presentation. This assumption is valid in
the weak-coupling regime which we are interested in for the qualitative
study. Also, we refer to the superconducting phase with $\Delta_{p}$
as $p$SC with $p=e$ or $o$ from now on.

The even-to-odd phase transition is determined by comparing the zero
temperature (Gibbs) free energies of eSC and oSC phases which are
calculated through
\begin{align}
{\cal F}_{p}(B_{z}) & =\frac{\Delta_{p}^{2}}{4g_{p}}+\check{\sum_{\boldsymbol{k}}}\frac{\text{Tr}[H_{0}(\boldsymbol{k})]-\sum_{n}E_{n}^{(p)}(B_{z},\boldsymbol{k})}{2}
\end{align}
with the positive energy $E_{n}^{(p)}(B_{z},\boldsymbol{k})$ of the
BdG Hamiltonian $H_{\text{BdG}}^{(p)}(B_{z},\boldsymbol{k})$. Fig.
\ref{fig:fig1}(a) illustrates the free energies of the normal, eSC,
and oSC phases from which the normal phase free energy in the zero-field
is subtracted. The parameters used are written in the caption of Fig.
\ref{fig:fig1}. The qualitative features of the system are well displayed
with this set of parameters. $g_{e}$ is chosen so that $\Delta_{e}=0.004$
is obtained by Eq. \eqref{eq:GapEq}, which is used throughout this
work unless otherwise noted.

Each curve in Fig. \ref{fig:fig1}(a) is well described by 
\begin{equation}
{\cal F}_{p}(B_{z})={\cal F}_{p}(0)-\frac{1}{2}\chi_{\text{spin}}^{(p)}B_{z}^{2},\label{eq:Free-energy}
\end{equation}
where the curvatures $\chi_{\text{spin}}^{(p)}$ of the curves are
understood as the spin susceptibility of the $p$SC state, while $\chi_{\text{spin}}^{(n)}$
denotes the normal phase spin susceptibility. The cross point at the
Pauli-limiting field $B_{z}=B_{z,\text{P}}$ between the normal (black
dashed line) and eSC (black line) phases marks the first order transition
between the normal and eSC phase. Using Eq. \eqref{eq:Free-energy},
$B_{z,\text{P}}$ is given by
\begin{equation}
B_{z,\text{P}}=\sqrt{\frac{{\cal F}_{n}(0)-{\cal F}_{e}(0)}{(\chi_{\text{spin}}^{(n)}-\chi_{\text{spin}}^{(e)})/2}}.
\end{equation}
Compared to the conventional Pauli-limiting critical field referred
to as the Chandrasekhar-Clogston field $B_{z,\text{CC}}=\sqrt{2\{{\cal F}_{n}(0)-{\cal F}_{e}(0)\}/\chi_{\text{spin}}^{(n)}}$,
$B_{z,\text{P}}$ is several times larger because of the non-vanishing
$\chi_{\text{spin}}^{(e)}$ due to the sizable Rashba spin-orbit couplings
\citep{Maruyama2012,Skurativska2021}.

The red, blue, and green lines denote the oSC free energies with $g_{o}=g_{e}$,
$g_{o}=1.17g_{e}$ and $g_{o}=1.2g_{e}$. Since $\chi_{\text{spin}}^{(o)}=\chi_{\text{spin}}^{(n)}$
as shown in Fig. \ref{fig:fig1}(a), the transition due to the Pauli
paramagnetic depairing does not occur between the normal phase and
the oSC state. The crossing point between the free energies of eSC
and oSC phases for a given $g_{o}$ indicates the even-to-odd transition
observed in the experiment \citep{Khim2021}. Moreover, the slope
of the free energies at the crossing point are different which means
the transition is of the first order and the magnetization changes
discontinuously at the transition.

Note that eSC state can be more stable than the oSC state at the zero-field
even if $g_{o}>g_{e}$, because the inter-layer spin-independent hoppings
$\varepsilon_{10}$ and $\varepsilon_{20}$ effectively weakens $g_{o}$.
The critical ratio $r_{c}\equiv g_{o,c}/g_{e}$, for which we obtain
$1.207$ for the aforementioned parameters, depends on the model parameters
$t,\;\mu,\;\alpha_{\text{R}},$ \emph{etc}. Above the critical ratio,
the oSC state is the superconducting ground state of the system at
the zero-field. In the two-dimensional limit in which the ratio $\alpha_{\text{R}}/\text{max}(|t_{c,1}|,|t_{c,2}|)$
is infinite, the electrons do not discern the trivial gap function
$\sigma_{0}$ from the sign-alternating gap function $\sigma_{z}$,
and thus $r_{c}\rightarrow1$.

Though the out-of-plane magnetic field is of our main interest, we
present the free energies under the in-plane magnetic field as well.
Fig. \ref{fig:fig1}(b) displays the free energies of the normal and
superconducting phases with in-plane magnetic fields $\boldsymbol{B}=B_{x}\hat{\boldsymbol{x}}$.
The free energies of the normal and the eSC phase cross at the Pauli-limiting
in-plane magnetic field $B_{x,\text{P}}$ which is smaller than $B_{z,\text{P}}$,
and this is consistent with the experiment \citep{Khim2021,Landaeta2022}.
When it comes to the oSC state, we do not see a first order transition
to the normal phase due to the Pauli depairing, while an exponential
decrease of the gap function is seen with the increasing in-plane
magnetic fields {[}See Appendix \ref{App:oSC-Bx} for details{]}.

Furthermore, it seems practically impossible to observe the first
order even-to-odd transition induced by the in-plane magnetic fields
even when $g_{o}/g_{e}$ is larger than $1$ unless sufficiently close
to the critical value. For example, when $g_{o}/g_{e}=1.2$, there
is a fair range of $B_{x}$ in which the oSC state is more stable
than the eSC state. However, for an intermediate ratio like $g_{o}/g_{e}=1.17$,
the distance between the free energies of the oSC phase and the normal
phase is very narrow when the free energies of the eSC and the oSC
states are comparable. Up to the impurity and finite-temperature effects,
$g_{o}/g_{e}\ge1$ is consistent with the experimental result where
a phase transition by the in-plane magnetic field is not identified
\citep{Khim2021}.

Thus far, we have demonstrated that the BdG model Hamiltonian with
the Zeeman term for the locally non-centrosymmetric system exhibits
a first order even-to-odd phase transition in the superconducting
phase under the out-of-plane magnetic fields at the zero-temperature.
An interesting point, not captured by the mean-field analysis, is
that first order transitions usually accompany hysteresis because
a system can still be in a metastable state. The range of the meta-stability
can be related with a collective mode \citep{LandauBook}, especially
the BS mode when a transition between two superconducting phases is
concerned, which is the main subject of this work.

\section{Bardasis-Schrieffer mode around the transition}

\subsection{Generalized Random Phase Approximation\protect\label{sec:GRPA}}

To study the BS mode, we use the generalized random phase approximation
(GRPA) \citep{Anderson1958,Rickayzen1959,Bardasis1961,Boyack2020},
which is one of the primary methods to incorporate the effect of the
collective modes in the superconducting phase. Before applying the
method to our case, we first briefly introduce the formulation of
the generalized random-phase approximation.

Concerned with the linear optical response of the fluctuation from
the subdominant pairing channels in a superconductor, we consider
an attractive electronic interaction consistent with the gap equation
in Eq. \eqref{eq:GapEq}:
\begin{equation}
\hat{V}=-\frac{1}{2}\sum_{p}\check{\sum_{k_{1},k_{2},q}}g_{p}\hat{\varPi}_{p}(k_{1},k_{1}-q)[\hat{\varPi}_{p}(k_{2},k_{2}-q)]^{\dagger},\label{eq:V}
\end{equation}
where $\hat{\varPi}_{p}(k_{1},k_{2})=\hat{\Psi}_{k_{1}}^{\dagger}\tau_{+}^{(p)}\hat{\Psi}_{k_{2}}$
with $\tau_{\pm}^{(p)}=(\tau_{x}^{(p)}\pm i\tau_{y}^{(p)})/2$. $\sum_{p}$
is the summation over the pairing channels labeled by $p=e$ and $p=o$.
While the pairing interaction $\hat{V}$ derived solely from the on-site
attractive interaction would have given us $g_{o}=g_{e}$, a general
pairing interaction gives us $g_{o}\neq g_{e}$. Thus we consider
the cases with $g_{o}\neq g_{e}$ as well as $g_{o}=g_{e}$. Other
pairing channels such as those discussed in Refs. \citep{Skurativska2021,Mockli2021}
do not couple linearly to the light because of the symmetries of $H_{0}$
with negligible $\lambda_{\text{I}}$ {[}See Appendix \ref{App:The-other-odd-parity}
for details{]}.

Adding $\hat{V}$ to the normal phase action ${\cal S}_{0}=\frac{1}{2}\check{\sum}_{k}\hat{\Psi}_{k}^{\dagger}(i\omega-H_{0})\hat{\Psi}_{k}$
and using the Hubbard-Stratonovich transformation, we obtain the following
total action for $p$SC phase with pairing fluctuations under the
external scalar and vector fields:
\begin{align}
{\cal S}= & {\cal S}_{e}^{(p)}+{\cal S}_{e-\eta}+S_{\text{\ensuremath{\eta}}}+{\cal S}_{e-A},\\
{\cal S}_{e}^{(p)}= & \frac{1}{2}\check{\sum_{k}}\hat{\Psi}^{\dagger}(k)\{-i\omega+H_{\text{BdG}}^{(p)}(\boldsymbol{k})\}\hat{\Psi}(k),\\
{\cal S}_{\eta}= & \frac{1}{2}\sum_{p,a}\check{\sum_{q}}\frac{\hat{\eta}_{a}^{(p)}(q)\hat{\eta}_{a}^{(p)}(-q)}{g_{p}},\label{eq:Seta}\\
{\cal S}_{e-\eta}= & \frac{1}{2}\sum_{p,a}\check{\sum_{k,q}}\hat{\Psi}^{\dagger}(k+q)\{\eta_{a}^{(p)}(q)\tau_{a}^{(p)}\}\hat{\Psi}(k),\label{eq:Se-eta}\\
{\cal S}_{e-A}= & \frac{1}{2}\check{\sum_{k,q}}\hat{\Psi}^{\dagger}(k+q)\{\Gamma_{1}(\boldsymbol{k},\boldsymbol{q})+\Gamma_{2}(k,q)\}\hat{\Psi}(k).
\end{align}
The auxiliary bosonic fields $\eta_{1}^{(p)}$ and $\eta_{2}^{(p)}$
represent the real and imaginary parts of the fluctuation in the pairing
channel $M_{p}$, respectively. They correspond to the amplitude and
the phase fluctuation, respectively, when $\Delta_{p}$ is real. $\Gamma_{1}$
and $\Gamma_{2}$ are the paramagnetic and diamagnetic light-matter
coupling vertices, respectively, and expressed as 
\begin{align}
\Gamma_{1}(\boldsymbol{k},q) & =\sum_{\mu=0}^{3}{\cal V}_{\mu}(\boldsymbol{k},\boldsymbol{q}){\cal A}_{\mu}(q),\\
\Gamma_{2}(k,q) & =\check{\sum_{k'}}\sum_{i,j=1}^{3}[m_{\boldsymbol{k}}^{-1}]_{ij}{\cal A}_{i}(q-k'){\cal A}_{j}(k'),
\end{align}
with the four-velocity operator ${\cal V}_{\mu}(\boldsymbol{k},\boldsymbol{q})=(2\tau_{z},\tau_{0}\partial_{i}\{H_{0}(\boldsymbol{k}+\boldsymbol{q})+H_{0}(\boldsymbol{k})\})/2$
and the inverse of the mass matrix $[m_{\boldsymbol{k}}^{-1}]_{ij}=\tau_{z}\partial_{i}\partial_{j}H_{0}(\boldsymbol{k})$.
Here, we define a four-potential $\mathcal{A}_{\mu}=|e|(-iA_{0},\boldsymbol{A})$
by multiplying the unit charge $|e|$ to the conventional four-potential
for conciseness.

The effective action for ${\cal A}_{\mu}$ and $\eta_{a}^{(p)}$ is
obtained by integrating out the fermionic degree of freedom $\hat{\Psi}$
and expanding the resultant action to the second order of ${\cal A}_{\mu}$
and $\eta_{a}^{(p)}$:
\begin{align}
{\cal S}_{\text{eff}} & =\frac{1}{2}\check{\sum_{q}}({\cal A_{\mu}},\eta_{a}^{(p)})_{\text{-}q}\varLambda(q)\bigg(\begin{matrix}{\cal A}_{\nu}\\
\eta_{b}^{(p')}
\end{matrix}\bigg)_{q},\label{eq:effectiveAction}
\end{align}
with 
\begin{equation}
\varLambda(q)=\bigg(\begin{matrix}K_{\mu\nu} & L_{\mu b}^{(p')}\\
R_{a\nu}^{(p)} & \frac{\delta_{pp'}\delta_{ab}}{g_{p}}+\Pi_{ab}^{(p,p')}
\end{matrix}\bigg)_{q},
\end{equation}
whose sub-blocks are given by
\begin{align}
K_{\mu\nu}(q)= & \frac{1}{2}\check{\sum_{k}}\text{Tr}[{\cal V}_{\mu}(\boldsymbol{k},\boldsymbol{q})G(k+q){\cal V}_{\nu}(\boldsymbol{k},\boldsymbol{q})G(k)]\nonumber \\
 & +\check{\sum_{k}}\text{Tr}[G(k)[m_{\boldsymbol{k}}^{-1}]_{\mu\nu}],\\
L_{\mu a}^{(p)}(q)= & \frac{1}{2}\check{\sum_{k}}\text{Tr}[{\cal V}_{\mu}(\boldsymbol{k},\boldsymbol{q})G(k+q)\tau_{a}^{(p)}G(k)],\\
R_{a\mu}^{(p)}(q)= & \frac{1}{2}\check{\sum_{k}}\text{Tr}[\tau_{a}^{(p)}G(k+q){\cal V}_{\mu}(\boldsymbol{k},\boldsymbol{q})G(k)],\\
\Pi_{ab}^{(p,p')}(q)= & \frac{1}{2}\check{\sum_{k}}\text{Tr}[\tau_{a}^{(p)}G(k+q)\tau_{b}^{(p')}G(k)].
\end{align}
Here, $[m_{\boldsymbol{k}}^{-1}]_{\mu\nu}$ is zero when either of
$\mu$ or $\nu$ is 0. The basic symmetry properties of the kernels
are $K_{\mu\nu}(q)=K_{\nu\mu}(-q)$, $\Pi_{ab}^{(p,p')}(q)=\Pi_{ba}^{(p',p)}(-q)$,
and $R_{a\mu}^{(p)}(q)=L_{\mu a}^{(p)}(-q)=[L_{\mu a}^{(p)}(q)]^{*}$.
The real-frequency kernels are obtained by the analytical continuation
$i\Omega\rightarrow\Omega^{+}=\Omega+i\epsilon$. $\epsilon=10^{-6}=2.5\times10^{-4}\Delta_{e}$
is used throughout this work unless otherwise noted.

Note that $H_{\text{BdG}}^{(e)}$ and $H_{\text{BdG}}^{(o)}$ are
invariant under the \emph{inversion }symmetry operators $\tau_{0}{\cal I}$
and $\tau_{z}{\cal I}$, respectively. These symmetries of the BdG
Hamiltonians provide a selection rule at $\boldsymbol{q}=\boldsymbol{0}$.
Considering the parities of the vertices $\mathcal{V}_{\mu}$ and
$\tau_{a}^{(M)}$ under the inversion symmetries of $H_{\text{BdG}}^{(p)}$,
several components of the kernel $\varLambda(i\Omega,\boldsymbol{0})$
are eliminated the kernel is reduced into two blocks:

\begin{align}
\varLambda(i\Omega)\sim & \biggg(\begin{matrix}K_{00} & L_{0b}^{(p)}\\
R_{a0}^{(p)} & \tilde{\Pi}_{ab}^{(p)}
\end{matrix}\biggg)\oplus\biggg(\begin{matrix}K_{ij} & L_{ib}^{(\bar{p})}\\
R_{aj}^{(\bar{p})} & \tilde{\Pi}_{ab}^{(\bar{p})}
\end{matrix}\biggg),\label{eq:reducedK}
\end{align}
where $\tilde{\Pi}_{ab}^{(p)}=g_{p}^{-1}\delta_{ab}+\Pi_{ab}^{(p,p)}$.
$\bar{p}$ denotes the odd-parity(even-parity) pairing channel in
eSC(oSC) state, which is the subdominant pairing channel. Eqs. \eqref{eq:effectiveAction}
and \eqref{eq:reducedK} explicitly show that the fluctuations of
the dominant pairing channel are coupled to the density-density response
$K_{00}$ in the first block in Eq. \eqref{eq:reducedK} whereas the
subdominant fluctuations are involved in the optical response in the
second block in Eq. \eqref{eq:effectiveAction} when $L_{ib}^{(\bar{p})}$
and $R_{aj}^{(\bar{p})}$ are finite. To discuss the linear optical
response of the fluctuation of the subdominant channel, we focus on
the second block in the remainder of our presentation.

\subsection{Bardasis-Schrieffer mode at the zero-field\label{subsec:BSmode_at_zero_field}}

\begin{figure}[t]
\centering\includegraphics[width=1\columnwidth]{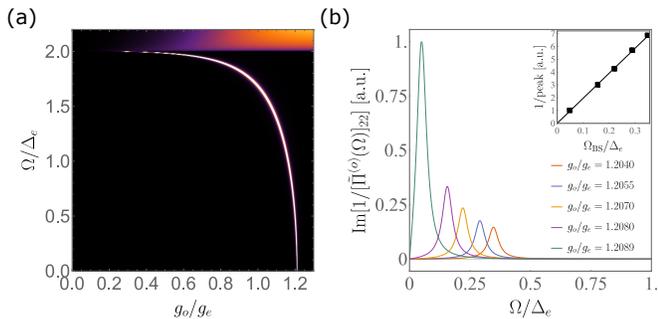}\caption{\label{fig:fig2}False color plot of $\text{Im}[1/[\tilde{\Pi}^{(o)}(\Omega)]_{22}]$
on the $g_{o}/g_{e}-\Omega$ plane in the eSC state. The energy $\Omega$
of the incident light is normalized by the magnitude $\Delta_{e}$
of the gap function in eSC state. (b) $\text{Im}[1/[\tilde{\Pi}^{(o)}(\Omega)]_{22}]$
for several $g_{o}/g_{e}$. For (b), $\epsilon=10^{-4}=2.5\times10^{-2}\Delta_{e}$
is exceptionally used. The inset plots the peak position in (b) and
the inverse of the peak height in (b).}
\end{figure}

Armed with the GRPA method, we study the BS mode originating from
the subdominant pairing fluctuation $\eta_{1}^{(\bar{p})}$ and $\eta_{2}^{(\bar{p})}$
\citep{Bardasis1961,Maiti2015,He2020,Boyack2020,Sauls2022}. The equation
of motion for the BS mode is given by $0=\delta{\cal S}_{\text{eff}}/\delta\eta_{a}^{(\bar{p})}$
which is rearranged into
\begin{equation}
\eta^{(\bar{p})}(q)=-[\tilde{\Pi}^{(\bar{p})}(q)]^{-1}R^{(\bar{p})}(q){\cal A}(q).\label{eq:LR-eta}
\end{equation}
Finding the singularity of the right-hand side (rhs) in Eq. \eqref{eq:LR-eta}
by solving $\det[\tilde{\Pi}^{(\bar{p})}(q)]=0$, the dispersion $\Omega_{\text{BS}}^{(\bar{p})}(\boldsymbol{q})$
of the BS is obtained. In general, $\Omega_{\text{BS}}^{(\bar{p})}(\boldsymbol{q})$
has its minimum at $\boldsymbol{q}=0$, and we refer to $\Omega_{\text{BS}}\equiv\Omega_{\text{BS}}^{(\bar{p})}(\boldsymbol{0})$
as the gap of BS mode.

In the eSC state, we show that $[\tilde{\Pi}^{(o)}(\Omega)]_{12}$
and $[\tilde{\Pi}^{(o)}(\Omega)]_{21}$ are vanishingly small, and
$[\tilde{\Pi}^{(o)}(\Omega)]_{11}$ is finite in Appendix. \ref{App:Pi_Omega_Bash-even}.
Thus, the zero of $\det[\tilde{\Pi}^{(o)}(\Omega)]\approx[\tilde{\Pi}^{(o)}(\Omega)]_{11}[\tilde{\Pi}^{(o)}(\Omega)]_{22}$
is largely identical to the zero of $[\tilde{\Pi}^{(o)}(\Omega)]_{22}$,
and the BS mode can be found by looking into the inverse of $[\tilde{\Pi}^{(o)}(\Omega)]_{22}$.
Figure \ref{fig:fig2}(a) shows $\text{Im}[1/[\tilde{\Pi}^{(o)}(\Omega)]_{22}]$
in the eSC state at the zero-field over varying $g_{o}/g_{e}$. The
gap of the BS mode $\Omega_{\text{BS}}$ is clearly identified. Increasing
$g_{o}/g_{e}$ drops $\Omega_{\text{BS}}$, and $\Omega_{\text{BS}}$
becomes zero at the critical ratio $r_{c}=g_{o,c}/g_{e}\sim1.21$
and disappears for larger $g_{e}/g_{o}$.

To relate the gap of the BS mode $\Omega_{\text{BS}}$ with the stability
of the superconducting ground state against the other pairing channel,
we first derive a semi-analytical expression for $\Omega_{\text{BS}}$
which is given by 
\begin{align}
\frac{2y\arcsin y}{\sqrt{1-y^{2}}}= & \frac{1}{g_{o}N_{\text{tot}}\langle\sin^{2}\chi\rangle_{\text{FS}}}-\frac{1}{g_{e}N_{\text{tot}}},\label{eq:OmegaBS_1}
\end{align}
where $y=\Omega_{\text{BS}}/2\Delta_{e}$, and $N_{\text{tot}}=N_{1}+N_{2}$
with $N_{i=1,2}$ being the density of states of the $i$-th Fermi
surface given by $\xi_{i}=0$. {[}The \emph{total} density of states
is $2N_{\text{tot}}$ due to the Kramers' degeneracy.{]} Here, $\langle\cdots\rangle_{\text{FS}}=\sum_{i=1}^{2}N_{i}\langle\cdots\rangle_{\text{FS},i}/(N_{1}+N_{2})$
with $\langle\cdots\rangle_{\text{FS},i}$ being the angular average
over the $i$-th Fermi surface {[}For the derivation, see Appendix
\ref{App:Pi_Omega_Bash-even}{]}.

The rhs of Eq. \eqref{eq:OmegaBS_1} can be related with the superconducting
phase transition temperatures if we note that, in the weak-coupling
limit, the phase transition temperatures for the eSC state and the
preempted oSC state are given by $g_{e}N_{\text{tot}}=-1/\ln(T_{c,e}/\Lambda)$
and $\langle\sin^{2}\chi\rangle g_{o}N_{\text{tot}}=-1/\ln(T_{c,o}/\Lambda)$,
respectively, where $\Lambda$ is a cutoff. Substituting these formulae
to Eq. \eqref{eq:OmegaBS_TceTco}, we obtain a simple relation between
$\Omega_{\text{BS}}$ and $T_{c,o}/T_{c,e}$: 
\begin{equation}
\frac{2y\arcsin y}{\sqrt{1-y^{2}}}=-\ln\frac{T_{c,o}}{T_{c,e}}.\label{eq:OmegaBS_TceTco}
\end{equation}
Note that a real solution $y$ exists as long as $T_{c,o}\le T_{c,e}$,
while it ceases to exist as soon as $T_{c,o}>T_{c,e}$. This implies
that $\Omega_{\text{BS}}=0$ is an indication of the phase transition
between two superconducting states.

Although Eq. \eqref{eq:OmegaBS_TceTco} is derived by assuming the
weak-coupling limit, let us make use of it to estimate $\Omega_{\text{BS}}$
in ${\rm CeRh_{2}As_{2}}$ at the zero-field. Adopting $T_{c,o}/T_{c,e}=0.87$
which is estimated for ${\rm CeRh_{2}As_{2}}$ in Ref. \citep{Khim2021},
we find $\Omega_{\text{BS}}\sim0.51\Delta_{e}$. Therefore, the BS
mode in ${\rm CeRh_{2}As_{2}}$ may be expected to exist below the
midst of the superconducting quasiparticle excitation gap, which is
a favorable condition to discern the signature of the BS mode from
the contributions from the quasiparticle excitations. It should be
stressed that this estimation of $\Omega_{\text{BS}}$ from Eq. \eqref{eq:OmegaBS_TceTco}
has nothing to do with our choice of the parameters such as $t,\;\mu,\;\alpha_{\text{R}}$
for the normal phase Hamiltonian; it is a model-independent result
under weak-coupling assumption.

In Fig. \ref{fig:fig2}(b), the intensity of the BS mode peaks in
$\text{Im}[1/[\tilde{\Pi}^{(o)}(\Omega)]_{22}]$ is drawn for several
$g_{o}/g_{e}$ around the critical point. As the critical point is
approached, the intensities of the peaks raise. The relation between
$\Omega_{\text{BS}}$ and the inverse of the peak height is shown
in the inset which clearly reveals that the peak height is proportional
to $\Omega_{\text{BS}}^{-1}$ for small $\Omega_{\text{BS}}$. $\text{Im}[1/[\tilde{\Pi}^{(e)}(\Omega)]_{22}]$
in the oSC state exhibits qualitatively identical trend except that
the BS mode appears when $g_{o}>g_{o,c}$.

It should be remarked that ${\rm peak}\propto\Omega_{\text{BS}}^{-1}$
is a consequence of generic properties of the linear response kernel
$\Pi^{(o)}(-\Omega)=[\Pi^{(o)}(\Omega)]^{\dagger}$ \citep{giuliani_vignale_2005}
which ensures that $\det\tilde{\Pi}^{(o)}(\Omega)$ is a real even
function of $\Omega$ for $\Omega<2\Delta_{e}$. Therefore, $\det\tilde{\Pi}^{(o)}(\Omega)\propto\Omega_{\text{BS}}^{2}-\Omega^{2}$
when $\Omega_{\text{BS}}$ and $\Omega$ are small. Provided that
$\det\tilde{\Pi}^{(o)}(\Omega)$ and $[\tilde{\Pi}^{(o)}(\Omega)]_{22}$
are proportional, we know $\text{Im}[1/[\tilde{\Pi}^{(o)}(\Omega)]_{22}]\propto\delta(\Omega_{\text{BS}}^{2}-\Omega^{2})\propto\delta(\Omega_{\text{BS}}-\Omega)/\Omega_{\text{BS}}$,
and thus the intensity of the BS mode increases as $\Omega_{\text{BS}}$
approaches to zero even when magnetic fields are applied. The same
argument applies to the BS mode in the oSC state.

\begin{figure*}
\centering\includegraphics[width=0.95\textwidth]{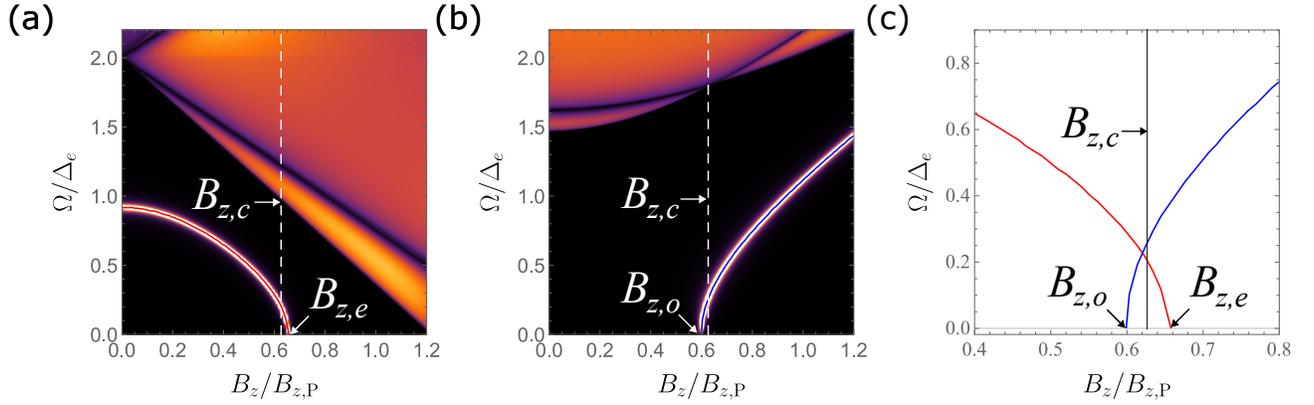}

\caption{\label{fig:fig3}(a) False color plot of $\text{Im}[1/[\tilde{\Pi}^{(o)}(\Omega)]_{22}]$
on the $B_{z}-\Omega$ plane in the eSC state when $g_{o}/g_{e}=1.17$.
(b) False color plot of $\text{Im}[[1/\tilde{\Pi}^{(e)}(\Omega)]_{22}]$
on the $B_{z}-\Omega$ plane in the oSC state when $g_{o}/g_{e}=1.17$.
The vertical axes of both figures are normalized by the magnitude
$\Delta_{e}$ of the gap function in eSC state, while the horizontal
axes are normalized by the Pauli magnetic field $B_{z,\text{P}}$
of the eSC state. The vertical dashed lines in (a) and (b) mark the
critical magnetic field at which the even-to-odd transition occurs.
The critical magnetic field is obtained from Fig. \ref{fig:fig1}(a).
For (b), $\Delta_{o}=0.001$ is used. A rather larger $\epsilon=10^{-4}$
in $\Omega^{+}$ is used exceptionally for (c) to avoid the numerical
issues concerning the extremely sharp peaks.}
\end{figure*}

\subsection{Bardasis-Schrieffer mode under $B_{z}$\label{subsec:BSmode_at_finite_field}}

Figures \ref{fig:fig3}(a) and \ref{fig:fig3}(b) show the imaginary
part of the inverse of the relevant component in $\tilde{\Pi}^{(\bar{p})}(\Omega^{+})$
in the $p$SC state for $\boldsymbol{B}=B_{z}\hat{\boldsymbol{z}}$.
Here, $g_{o}=1.17g_{e}$ is used to make the features of figures easily
recognizable. The reddish region of each figure represents the BdG
quasiparticle excitations. Below the quasiparticle continuum, the
curves corresponding to the gap of the BS mode $\Omega_{\text{BS}}(B_{z})$
in each $p$SC is clearly depicted. The red and blue lines are drawn
over the curves for a guide to the eye. The vertical dashed lines
in both figures denote the even-to-odd critical fields $B_{z,c}$
identified in Fig. \ref{fig:fig1}(a).

It is clearly seen that $\Omega_{\text{BS}}(B_{z})$ in eSC (oSC)
phase is lowered as the external magnetic fields increase (decreases).
This is consistent with the behavior of the BdG quasiparticle excitation
gap. Also, it has to be noted that $\Omega_{\text{BS}}(B_{z,c})$
is finite, while it becomes zero at $B_{z}=B_{z,e}>B_{z,c}$ and $B_{z}=B_{z,o}<B_{z,c}$
in eSC and oSC phases, respectively. Recalling that the eSC and oSC
phases are the equilibrium ground states in $B_{z}<B_{z,c}$ and $B_{z}>B_{z,c}$,
respectively, the softening of those collective modes occurs outside
the thermodynamic equilibrium \citep{LandauBook}. Understood as a
precursor of an instability of a state, $B_{z,e}$($B_{z,o}$) could
be understood as the boundary to which the eSC(oSC) state can persist
to exist as a metastable state. Therefore, if the experimentally observed
hysteresis \citep{Khim2021} may originate from the metastable eSC
and oSC states, it is expected that the BS mode is almost gapless
at the boundaries of the hysteresis curve.

As explained before, the peak height of the BS mode increases as the
gap of the BS mode decreases.

\section{Linear optical response in LNCS superconductor}

Thus far, we have demonstrated that the gap of the BS mode from the
pairing channel with opposite parity to the ground state is lowered
near the even-to-odd phase transition and becomes gapless at a critical
point which may be identified with a boundary of the hysteresis. In
this section, we study the linear optical response from the BS mode.

The linear optical response incorporating the effect of the sub-dominant
pairing fluctuation is derived from ${\cal J}_{i}(q)/|e|=\delta S_{\text{eff}}/\delta{\cal A}_{i}(-q)$
with $S_{\text{eff}}$ in Eq. \eqref{eq:effectiveAction}:
\begin{align}
{\cal J}_{i}(q)/|e|= & K_{ij}(q){\cal A}_{j}(q)+L_{ia}^{(\bar{p})}(q)\eta_{a}(q).\label{eq:LR-J}
\end{align}
Substituting Eq. \eqref{eq:LR-eta} into Eq. \eqref{eq:LR-J}, we
have ${\cal J}_{i}(q)/|e|=\tilde{K}_{ij}(q){\cal A}_{j}(q)$ with
\begin{align}
\tilde{K}(q) & =K(q)-L^{(\bar{p})}(q)[\tilde{\Pi}^{(\bar{p})}]^{-1}R^{(\bar{p})}(q),\label{eq:tildeK}
\end{align}
where the second term in the rhs of Eq. \eqref{eq:tildeK} includes
the contribution from the BS mode. Therefore, the BS mode would be
detected as a peak in the optical absorption spectrum unless $L_{ia}^{(\bar{p})}(\Omega)$
and $R_{aj}^{(\bar{p})}(\Omega)$ are zero.

\subsection{Multiband-assisted optical transition\protect\\and non-zero $L^{(o)}$
and $R^{(o)}$\label{subsec:Multiband-assisted}}

\begin{figure*}
\centering\includegraphics[width=0.95\textwidth]{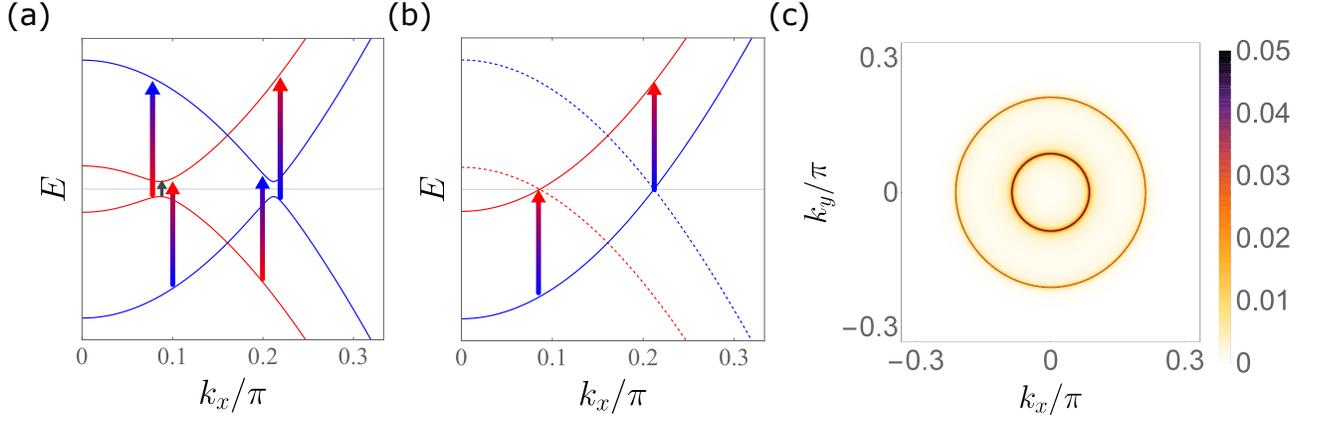}\caption{\label{fig:fig4}The band structure of (a) $H_{\text{BdG}}$ in Eq.
\eqref{eq:CeRhAs-BdG} and (b) $H_{0}$ in Eq. \eqref{eq:CeRhAs-normal}.
The gray arrows in (a) correspond to the forbidden optical transitions
due to the vanishing matrix elements of the velocity operator ${\cal V}_{z}$.
The arrows with color gradient indicate the optical transition making
$L_{z,2}^{(o)}$ and $R_{2,z}^{(o)}$ finite. $\Delta_{e}=0.05$ are
used for (a) and (b). (c) False color plot for the integrand $F(\boldsymbol{k},0)$
of $L_{z,2}^{(o)}$ with $\Delta_{e}=0.004$.}
\end{figure*}

$L_{ia}^{(\bar{p})}(\Omega)$ and $R_{aj}^{(\bar{p})}(\Omega)$ are
frequently overlooked in literature. A partial reason for this may
come from the conventional wisdom that the matrix elements of the
velocity operators vanishes in the BCS model for the conventional
superconductivity with a single electronic band \citep{Ahn2021} as
we show explicitly below.

Unlike the conventional models for the trivial superconductivity with
a single band normal phase Hamiltonian, however, the presence of multiple
electronic bands can render $L_{i,a}^{(\bar{p})}(\Omega)$ and $R_{a,j}^{(\bar{p})}(\Omega)$
finite \citep{Kamatani2022}. To emphasize the role of the multiple
electronic bands, we provide a detailed analysis of $L_{ia}^{(o)}(\Omega)$
and $R_{aj}^{(o)}(\Omega)$ in this section based on the analytical
expressions of them in the eSC state.

As shown in Appendix \ref{App:The-other-odd-parity}, the spatial
symmetries of $H_{0}$ allow $L_{z,2}^{(o)}$ to be finite whereas
$L_{x,2}^{(o)}$ and $L_{y,2}^{(o)}$ are forbidden. Also, as $L_{z,2}^{(o)}(\Omega)=[R_{2,z}^{(o)}(\Omega)]^{*}$,
we focus on $L_{z,2}^{(o)}(\Omega)$. The spectral representation
of $L_{z,2}^{(o)}$ is written down as 
\begin{equation}
L_{z,2}^{(o)}(\Omega)=\frac{1}{2}\check{\sum_{\boldsymbol{k}}}\sum_{m,n}\frac{\langle m|{\cal V}_{z}|n\rangle\langle n|\tau_{y}\sigma_{z}|m\rangle}{\Omega^{+}-E_{m}+E_{n}}\Theta_{mn},\label{eq:L_spectral}
\end{equation}
where the momentum dependence of the eigenstate $|m\rangle$ and energy
$E_{m}$ are omitted. $\Theta_{mn}\equiv\Theta(E_{m})-\Theta(E_{n})$
with the Heaviside step function $\Theta(x)$. The eigenenergies of
$H_{\text{BdG}}^{(e)}$ given by $E_{c(v),i}=\pm\sqrt{\xi_{i}^{2}+\Delta_{e}^{2}}$
for $i=1,2$. The corresponding eigenvectors are 
\begin{equation}
\begin{split}|c,i,\alpha(\beta)\rangle & =\left(\begin{matrix}\cos\frac{\Xi_{i}}{2}|\xi_{i},\alpha(\beta)\rangle\\
\sin\frac{\Xi_{i}}{2}|\xi_{i},\alpha(\beta)\rangle
\end{matrix}\right),\\
|v,i,\alpha(\beta)\rangle & =\left(\begin{matrix}-\sin\frac{\Xi_{i}}{2}|\xi_{i},\alpha(\beta)\rangle\\
\cos\frac{\Xi_{i}}{2}|\xi_{i},\alpha(\beta)\rangle
\end{matrix}\right),
\end{split}
\label{eq:CeRhAs_BdG_Eigenvector}
\end{equation}
with $e^{i\Xi_{i}}=(\xi_{i}+i\Delta_{e})/E_{c,i}$.

Let us first evaluate the matrix elements of the velocity operator
${\cal V}_{z}$ $\langle m|{\cal V}_{z}|n\rangle$ using Eqs. \eqref{eq:CeRhAs_BdG_Eigenvector},
\eqref{eq:CeRhAs_Eigenvectors1}, and \eqref{eq:CeRhAs_Eigenvectors2}.
The elements $\langle m|{\cal V}_{z}|n\rangle$ relevant to calculating
$L_{z,2}^{(o)}$ at the zero-temperature are given by 
\begin{align}
\langle c,i,\alpha|{\cal V}_{z}|v,j,\alpha'\rangle= & \sin\frac{\Xi_{i}-\Xi_{j}}{2}\langle i,\alpha|\partial_{z}H_{0}|j,\alpha'\rangle.\label{eq:J_v1c2_main}
\end{align}
Note that the rhs is zero when $i=j$. These elements correspond to
the forbidden transitions $\langle c,i,\alpha|{\cal V}_{z}|v,i,\alpha\rangle$
that are marked by gray arrows in Figure \ref{fig:fig4}(a), where
the energy bands of the BdG quasiparticles are drawn. Furthermore,
an explicit calculation using the eigenvectors in Eqs. \eqref{eq:CeRhAs_Eigenvectors1}
and \eqref{eq:CeRhAs_Eigenvectors2} shows that $\langle i,\alpha|\partial_{z}H_{0}|j,\alpha'\rangle\propto\delta_{\alpha\alpha'}$.
Therefore, only $\langle c,1,\alpha|{\cal V}_{z}|v,2,\alpha\rangle$
and the other elements related to this by the complex conjugation
or a replacement $1\leftrightarrow2$ or $\alpha\leftrightarrow\beta$
are finite. The arrows with color gradient in Fig. \ref{fig:fig4}(a)
represent the transitions related to these finite elements of the
velocity operator. Comparing it to the electronic band structure in
the normal phase displayed in Fig. \ref{fig:fig4}(b), these finite
transitions can be understood as the remnants of the interband transitions
in the normal phase which are marked by arrows in Fig. \ref{fig:fig4}(b).

To calculate $L_{z,2}^{(o)}(\Omega)$, we further need to evaluate
$\langle c,2,\alpha|\tau_{y}\sigma_{z}|v,1,\alpha\rangle$ which is
given by 
\begin{align}
\langle c,2,\alpha|\tau_{y}\sigma_{z}|v,1,\alpha\rangle & =\frac{e^{i(\zeta+\phi)}t_{\boldsymbol{k}}\cos\chi}{2i}\cos\frac{\Xi_{2}-\Xi_{1}}{2},\label{eq:c2_tau_y_sigma_z_v1_main}
\end{align}
and the other elements related to it by the complex conjugation or
the replacement $1\leftrightarrow2$ are also finite.

Substituting Eqs. \eqref{eq:J_v1c2_main} and \eqref{eq:c2_tau_y_sigma_z_v1_main}
into Eq. \eqref{eq:L_spectral} results in 
\begin{align}
L_{z,2}^{(o)}(\Omega)= & -\int F(\boldsymbol{k},\Omega)\text{d}^{d}\boldsymbol{k}/(2\pi)^{d},\\
F(\boldsymbol{k},\Omega)= & \frac{(E_{c,1}+E_{c,2})t(\boldsymbol{k})\cos\chi\sin(\Xi_{2}-\Xi_{1})}{(E_{c,1}+E_{c,2})^{2}-(\Omega^{+})^{2}}.\label{eq:F(k,Omega)}
\end{align}
Note that the possible singularities of $F(\boldsymbol{k},\Omega)$
are located at $|\Omega|=\Delta_{e}+|\xi_{1,\boldsymbol{k}}-\xi_{2,\boldsymbol{k}}|$
which are fairly distant from the region of interest $|\Omega|<2\Delta_{e}$.
Hence, it is a good approximation to set $\Omega=0$ in Eq. \eqref{eq:F(k,Omega)}.
We draw $F(\boldsymbol{k},0)$ in Fig. \ref{fig:fig4}(c). $F(\boldsymbol{k},0)$
has narrow positive peaks around the Fermi surfaces like the integrands
that are commonly encountered in the weak-coupling theory of superconductivity.
Around each Fermi surface, $F(\boldsymbol{k},0)\approx\frac{\Delta_{e}\cos^{2}\chi}{4E_{i}(\boldsymbol{k})}$,
and thus
\begin{equation}
L_{z,2}^{(o)}(\Omega)\approx-\check{\sum_{k}}\frac{\Delta_{e}\cos^{2}\chi}{4E_{i}}=-\frac{\Delta_{e}\langle\cos^{2}\chi\rangle_{\text{FS}}}{2g_{e}}.\label{eq:Loze}
\end{equation}
where we use the BCS gap equation $1/g_{e}=(N_{1}+N_{2})\int\text{d}\xi(\xi^{2}+\Delta_{e}^{2})^{-1/2}$
with $N_{i}$ the density of states of the $i$-th Fermi surface given
by $\xi_{i}=0$. Though $\Delta_{e}/g_{e}$ is small, $L_{z,2}^{(o)}(\Omega)$
is finite as long as $\langle\cos^{2}\chi\rangle_{\text{FS}}\neq0$,
which is proportional to the square of interlayer hopping $t_{c}^{2}$
. Also, we can know that $L_{z,2}^{(o)}(\Omega)$ decreases as $\alpha_{\text{R}}^{2}$
increases from Eq. \eqref{eq:Loze} as $N_{\text{tot}}$ converges
to a constant proportional $1/t$. For the estimation of $L_{z,2}^{(o)}$
under the weak-coupling assumption with $g_{o}=g_{e}$, see Appendix.
\ref{App:Estimation_Lz2}.

\subsection{Optical response under $B_{z}$}

\begin{figure*}[t]
\centering\includegraphics[width=0.95\textwidth]{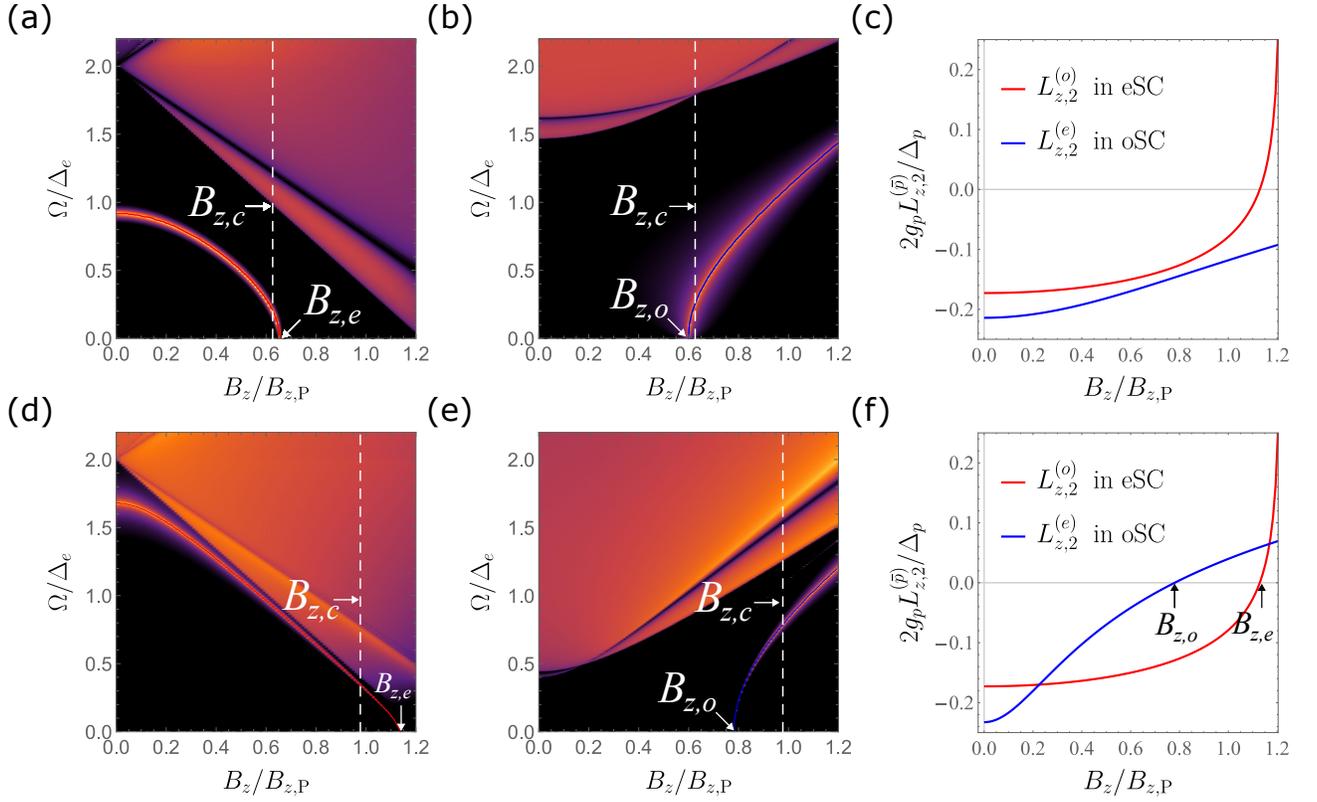}\caption{\label{fig:fig56}False color plots for $\text{Im}[\tilde{K}_{zz}(\Omega)]$
on the $B_{z}-\Omega$ plane. For (a), (b), and (c), $g_{o}/g_{e}=1.17$
is used, while (d), (e) and (f) are calculated with $g_{o}/g_{e}=1$.
(a) and (d) $\text{Im}[\tilde{K}_{zz}(\Omega)]$ in the eSC state,
(b) and (e) $\text{Im}[\tilde{K}_{zz}(\Omega)]$ in the oSC state.
(c) and (f) $L_{z,2}^{(\bar{p})}(\Omega)$ of both phases normalized
by $g_{p}/2\Delta_{p}$. The red and blue lines correspond to $L_{z,2}^{(o)}(\Omega)$
in the eSC state and $L_{z,2}^{(e)}(\Omega)$ in the oSC state, respectively.}
\end{figure*}

Provided that the fluctuation in the subdominant pairing channel is
linearly coupled to the light, the behavior of the gaps of the BS
modes around the even-to-odd transition can be investigated through
an optical measurement in the linear response regime. Figures \ref{fig:fig56}(a)
and \ref{fig:fig56}(b) show the imaginary part of the linear response
kernel $\tilde{K}_{zz}(\Omega)$ in Eq. \eqref{eq:tildeK} for varying
$B_{z}$ in the eSC and the oSC states, respectively, with $g_{o}=1.17g_{e}$.
It is easy to see the signature from the collective modes appearing
in $\text{Im}[1/\tilde{\Pi}_{22}^{(\bar{p})}(\Omega)]$ in Figs. \ref{fig:fig3}(a)
and \ref{fig:fig3}(b). Also, the BS mode is well separated from the
BdG quasiparticle pair-breaking continuum. Since $L_{z,a}^{(\bar{p})}(\Omega)$
is finite far below the pair-breaking continuum, it lets the collective
modes make distinguished contributions to $\text{Im}[\tilde{K}_{zz}(\Omega)]$.
At the point where the gap of the BS mode vanishes, the peak intensity
of the BS modes diverges and $\text{Im}[\tilde{K}_{zz}(\Omega)]$
exhibits the strongest peak from the BS mode. Thus, the absorption
peak in the optical response measurement is expected to be strongest
at the boundaries of the hysteresis curve around the even-to-odd transition.

Figs. \ref{fig:fig56}(d)\textasciitilde (f) shows ${\rm Im}[\tilde{K}_{zz}(\Omega)]$
and $L_{z,2}^{(\bar{p})}$ when $g_{o}=g_{e}$. Unlike Fig. \ref{fig:fig56}(a),
Fig. \ref{fig:fig56}(d) is featured by that the pair-breaking continuum
is around the $\Omega=0$ line when $\Omega_{\text{BS}}\sim0$. Because
of the lowered pair-breaking continuum, the magnitude of $L_{z,2}^{(o)}(\Omega)$
is diminished as $B_{z}$ increases. Fig. \ref{fig:fig56}(f) explicitly
shows how $L_{z,2}^{(o)}(\Omega)$ changes with $B_{z}$. For magnetic
fields close to either of $B_{z,e}$ and $B_{z,o}$ at which $\Omega_{\text{BS}}$
vanishes, the magnitude of $L_{z,2}^{(\bar{p})}(\Omega)$ rapidly
drops off. The diminished $L_{z,a}^{(\bar{p})}(\Omega)$ enfeebles
the intensity of the peak at $\Omega=\Omega_{\text{BS}}$. Especially
when $g_{o}=g_{e}$, the magnetic field $B_{z}$ rendering $L_{z,2}^{(\bar{p})}(0)=0$
coincides with the magnetic field at which $\Omega_{\text{BS}}=0$
occurs. Subsequently, the gapless BS mode from the $\bar{p}$-parity
pairing channel in $p$SC seems to have no effect on the linear optical
response in this limit because the coupling $L_{z,2}^{(\bar{p})}$
between the BS mode and the light vanishes.

However, this exact coincidence of $\Omega_{\text{BS}}=0$ and $L_{z,a}^{(\bar{p})}=0$
under the out-of-plane magnetic field is an unavoidable consequence
of symmetry. It turns out that the coincidence happens to occur because
we ignore the Ising-type spin-orbit coupling $\lambda_{\text{I}}$,
which results in $F_{A}(\boldsymbol{k})\equiv[\sigma_{z},H_{0}(\boldsymbol{k})]/4i={\cal V}_{z}(\boldsymbol{k})$.
Without $\lambda_{\text{I}}$, the zero-frequency response kernel
$\tilde{\Pi}_{22}^{(\bar{p})}$ is given by 
\begin{align}
\tilde{\Pi}_{22}^{(\bar{p})}(0)= & \frac{1}{g_{o}}-\frac{1}{g_{e}}-\frac{\check{\sum_{k}}\text{Tr}[\tau_{y}^{(o)}G(k)F_{A}(\boldsymbol{k})G(k)]}{\Delta_{p}}\\
= & \frac{1}{g_{o}}-\frac{1}{g_{e}}-\frac{2L_{z,2}^{(\bar{p})}(0)}{\Delta_{p}}.\label{eq:Origin_of_Artifact}
\end{align}
where we use $F_{A}(\boldsymbol{k})={\cal V}_{z}(\boldsymbol{k})$
in the second line. Given $g_{o}=g_{e}$, the rhs is zero when $L_{z,2}^{(\bar{p})}(0)=0$.
Adding the Ising-type spin-orbit coupling forces $F_{A}(\boldsymbol{k})\neq{\cal V}_{z}(\boldsymbol{k})$
and enables the gapless BS mode contribute to the linear optical response
in principle.

\section{Summary and Discussion}

We have investigated the BS modes from an odd- and even-parity pairing
channels in the eSC and oSC state, respectively, by using a generic
model for locally non-centrosymmetric superconductors involving two
orbital degrees of freedom. Our result based on the GRPA shows the
gap of the BS mode in the eSC (oSC) state is lowered with the increasing
(decreasing) out-of-plane magnetic field and eventually becomes gapless.
Since the softening of the BS modes is the precursor of the end of
the metastability of a superconducting state, we deduce that the softening
should occur at the boundaries of the hysteresis curve around the
first order even--to-odd transition.

As the BS modes considered in this work originate from the pairing
channels with the parity opposite from that of the ground state pairing,
there can be a finite linear coupling between the light and the BS
modes. We have demonstrated that the linear coupling is indeed finite
due to the presence of the multiple electronic bands, which can be
thought of as an intrinsic characteristic of a locally non-centrosymmetric
system with two orbital degrees of freedom in the primitive cell.
Therefore, we look forward that the signature of the collective mode
can be observed by measuring the linear optical response, especially
in the microwave regime, of ${\rm CeRh_{2}As_{2}}$ for which $\Omega_{\text{BS}}\sim0.51\Delta_{e}$
is expected at the zero-field.

It should be stressed that the detection the BS modes via an \emph{linear}
optical response measurement is a smoking gun signature from the bulk
of ${\rm CeRh_{2}As_{2}}$ evidencing the competing odd-parity pairing
channel. This is because the linear optical coupling is possible only
when the dominant and the sub-dominant pairing channels are opposite
in parity. Moreover, as exposed in Appendix \ref{App:The-other-odd-parity},
the light selectively couples to a particular set of odd-parity pairings.
Put into the group theoretical jargon, only the pairing channels belonging
to the irreducible representations of $\boldsymbol{J}_{i}$ are able
to affect the optical response in the linear response regime. Therefore,
the detection of the BS modes not only can be taken as a compelling
proof, i.e. sufficient, for the existence of the odd-parity pairing
channel, but also can place restrictions on the form of the odd-parity
pairing channels. It also deserves to be noted that the gap of the
BS mode in the oSC increases with increasing out-of-plane magnetic
field. This feature may be regarded as a proof of the parity-switching
at the first order transition in the superconducting phase of ${\rm CeRh_{2}As_{2}}$
because the gap of the BS mode should decreases if it were not for
the parity-switching.

Though the Pauli paramagnetic depairing is considered as the primary
cause of the first order transition in the superconducting state of
${\rm CeRh_{2}As_{2}}$, our findings and argument are applicable
to any superconducting systems exhibiting parity-switching transitions
between two superconducting states regardless of the underlying mechanism
and the transition order. An interesting application is the superconductivity
in a system hosting a structural instability \citep{Kozii2015,Venderbos2016,Wang2017,Wang2016}
e.g. ferroelectric instability. We address the cases in the two perspectives.
Firstly, if the even-to-odd transition is realized within the centrosymmetric
state of this system, it is possible to have a soft BS mode at the
transition, which may also be observed in an optical response measurement
in the linear response regime. Also, by noting that the topological
characterization of the superconductor can accompany the transition,
both of the fermionic and the collective excitations are gapless at
the transition and thus an intriguing phenomena such as the non-Fermi
liquid state could be brought about.

The second case is when such a transition occurs in the non-centrosymmetric
state. In this case, the superconducting phase could host an intriguing
topological phase transition between an even-parity dominant trivial
superconductivity and an odd-parity dominant topological superconductivity,
and a low-lying Leggett mode could appear at the transition \citep{Wang2017,Wang2016}.
The existence of such a topological phase transition implies there
are at least two competing pairing channels whose parities were opposite
if it were not for the inversion-breaking order. However, the inversion-breaking
order blurs the the sharp distinction between even- and odd-parity
pairings, which could lead both pairing channels to belonging to the
same irreducible representation of the symmetry group of the state.
In such a case, the BS mode from the competing pairing channel will
turn into a Leggett mode, which is discussed in Refs \citep{Wang2016,Wang2017}.
This Leggett mode can also be coupled linearly to the light due to
the absence of the inversion symmetry \citep{Kamatani2022}.

Lastly, a recent experiments suggests that the possibility of an inversion-breaking
antiferromagnetic order coexisting with superconductivity in ${\rm CeRh_{2}As_{2}}$
\citep{Kibune2022,Landaeta2022}. As the presence of the antiferromagnetic
order can reduce the group of the symmetries of the system, its potential
effect on the existence of BS modes and influence on the optical measurement
calls for further investigation.
\begin{acknowledgments}
We express our sincere thanks to Nico A. Hackner and P. M. R. Brydon
for generously sharing their unpublished preprint on the BS mode \citep{Hackner2022}.
We also thank Daniel Agterberg, Hongki Min, Yunsu Jang, Jiho Jang,
and Sungmo Kang for helpful discussions. S.B.C. was supported by the
National Research Foundation of Korea (NRF) grants funded by the Korea
government (MSIT) (2020R1A2C1007554) and the Ministry of Education
(2018R1A6A1A06024977).
\end{acknowledgments}

\appendix

\section{The other odd-parity channels\label{App:The-other-odd-parity}}

\subsection{When the ground state at the zero-field is an even-parity state}

In this section, we discuss the linear coupling between the other
odd parity channels and the light (or the current). We first note
that the presence of an inversion ${\cal I}=\sigma_{x}$ and a time-reversal
symmetry ${\cal T}=is_{y}{\cal K}$ enforces for the normal phase
Hamiltonian to take the following form
\begin{align}
H_{0}(\boldsymbol{k})= & \varepsilon_{00}(\boldsymbol{k})\sigma_{0}s_{0}+(\varepsilon_{10}(\boldsymbol{k})\sigma_{x}+\varepsilon_{20}(\boldsymbol{k})\sigma_{y})s_{0}\nonumber \\
 & +\sigma_{z}(\varepsilon_{31}(\boldsymbol{k})s_{x}+\varepsilon_{32}(\boldsymbol{k})s_{y}+\varepsilon_{33}(\boldsymbol{k})s_{z})
\end{align}
where $\varepsilon_{00}(\boldsymbol{k})$ and $\varepsilon_{10}(\boldsymbol{k})$
are even functions under $\boldsymbol{k}\rightarrow-\boldsymbol{k}$
while $\varepsilon_{20}(\boldsymbol{k})$ and $\varepsilon_{3i}(\boldsymbol{k})$
are odd functions. The linear coupling between a pairing channel and
the light is possible only when a pairing channel transforms like
one of the current operators $J_{i}$ under the symmetries of $H_{0}(\boldsymbol{k})$.
For ${\rm CeRh_{2}As_{2}}$, the point group $D_{4h}$ is the symmetry
of the Hamiltonian at $\Gamma$ in the Brillouin zone. By using the
symmetries of the point group $D_{4h}$, we analyze the selection
rule for odd-parity channels transforming like either of $k_{x}s_{y}-k_{y}s_{x}$,
$k_{z}s_{z}$, $k_{z}\sigma_{x}s_{z}$ and $k_{x}s_{x}+k_{y}s_{y}$
which are discussed in Refs. \citep{Skurativska2021}.

Table \ref{tab:CharacterTable} summarizes the parities of the current
operators and the form factors of those odd-parity channels under
several two-fold transformations. The signs tell whether $TO(\boldsymbol{k})T^{-1}=+O(T\boldsymbol{k})$
or $TO(\boldsymbol{k})T^{-1}=-O(T\boldsymbol{k})$ where $O$ represents
one of the currents or the form factors in the first column of the
Table \ref{tab:CharacterTable} and $T$ represents a symmetry transformation
in the first row of the table.

Firstly, the linear coupling between the in-plane currents $J_{x}$
and $J_{y}$ and the odd-parity gap functions in Table \ref{tab:CharacterTable}
is forbidden by, for example, $C_{2z}$. It is easy to see that the
odd-party channel transforming like $k_{x}s_{x}+k_{y}s_{y}$ or $k_{z}s_{z}$
can not be linearly coupled to the light because of $C_{2z}$ and
$C_{2x}$. The odd-parity channel labeled by $k_{x}s_{y}-k_{y}s_{x}$
transforms like $J_{z}$ for all two-fold symmetries in $D_{4h}$.
Indeed, $J_{z}$ and $k_{x}s_{y}-k_{y}s_{x}$ belong to the same irreducible
representation, and thus $k_{x}s_{y}-k_{y}s_{x}$ can be coupled to
the light as $\sigma_{z}$ can.

\begin{table}[t]
\begin{tabular}{|c|c|c|c|c|}
\hline 
 & ${\cal I}\;(\sigma_{x})$ & $C_{2z}\;(s_{z})$ & $C_{2x}\;(\sigma_{x}s_{x})$ & ${\cal A}\;(\sigma_{y}s_{x}{\cal K})$\tabularnewline
\hline 
\hline 
$J_{z}$, $\sigma_{z}$ & $-$ & $+$ & $-$ & $-$\tabularnewline
\hline 
\{$J_{x}$, $J_{y}$\} & $-$ & $-$ & $\{+,-\}$ & $-$\tabularnewline
\hline 
$k_{x}s_{y}-k_{y}s_{x}$ & $-$ & $+$ & $-$ & $+$\tabularnewline
\hline 
$k_{z}$$s_{z}$ & $-$ & $+$ & $+$ & $-$\tabularnewline
\hline 
$k_{x}s_{x}+k_{y}s_{y}$, $k_{z}\sigma_{x}s_{z}$ & $-$ & $+$ & $+$ & $+$\tabularnewline
\hline 
\end{tabular}\caption{\label{tab:CharacterTable}Character table for some two-fold and $M_{(abc)}$
is a mirror operation against a plane perpendicular to the vector
$(a,b,c)$. $C_{(110)}$ is a two-fold rotation around the axis $(1,1,0)$.
Here, $\boldsymbol{\Omega}=\hat{\boldsymbol{n}}_{110}\cdot\boldsymbol{s}.$}
\end{table}

In the above symmetry-based analysis, however, the details of the
electronic structure is not taken into consideration. For ${\rm CeRh_{2}As_{2}}$,
the large contribution to $\varepsilon_{33}(\boldsymbol{k})$ may
be supposed to originate from the Ising-type spin-orbit couplings
between next-nearest-neighboring ${\rm Ce}$ atoms. As long as this
Ising-type spin-orbit coupling is so negligible that $\varepsilon_{33}$
is also negligible compared to other $\varepsilon_{ij}$, we can show
that the coupling between $k_{x}s_{y}-k_{y}s_{x}$ and $J_{z}$ is
much smaller than that between $\sigma_{z}$ and $J_{z}$.

To prove it, we first note that the non-trivial part of the normal
phase Hamiltonian $\tilde{H}_{0}(\boldsymbol{k})\equiv H_{0}(\boldsymbol{k})-\varepsilon_{00}(\boldsymbol{k})\sigma_{0}s_{0}$
possesses an additional \emph{antiunitary} \emph{antisymmetry} ${\cal A}=U_{{\cal A}}{\cal K}$
of $\tilde{H}_{0}(\boldsymbol{k})$ with $U_{{\cal A}}=i\sigma_{y}s_{x}$.
It transforms under ${\cal A}$ as $U_{{\cal A}}\tilde{H}_{0}(\boldsymbol{k})^{*}U_{{\cal A}}^{\dagger}=-\tilde{H}(\boldsymbol{k})$.
By ${\cal A}$, the eigenvectors $|\xi_{1},\alpha\rangle$ and $|\xi_{1},\beta\rangle$
are related to $|\xi_{2},\alpha\rangle$ an $|\xi_{2},\beta\rangle$:
\begin{align}
{\cal A}|\xi_{1},\alpha\rangle & =\sum_{\alpha'=\alpha,\beta}[\Gamma_{{\cal A}}]_{\alpha',\alpha}|\xi_{2},\alpha'\rangle,\\
{\cal A}|\xi_{2},\alpha\rangle & =\sum_{\alpha'=\alpha,\beta}[-\Gamma_{{\cal A}}^{T}]_{\alpha',\alpha}|\xi_{1},\alpha'\rangle,
\end{align}
where $\Gamma_{{\cal A}}$ is a $2\times2$ unitary matrix. Here,
we use $U_{{\cal A}}=-U_{{\cal A}}^{T}$.

The antiunitary antisymmetry of $\tilde{H}_{0}$ is especially useful
when the linear coupling is computed between the current operator
$J_{z}$ and the pairing fluctuations with the form factor $M_{\boldsymbol{k}}$
in the eSC state with the trivial ground state gap function. In the
calculation, we frequently encounter terms such as ${\cal I}_{m\bar{m}}=\sum_{\alpha,\beta}\langle m,\alpha|J_{z}|\bar{m},\beta\rangle\langle\bar{m},\beta|M_{\boldsymbol{k}}|m,\alpha\rangle$
with $\bar{m}=-m$ being $1$ or $2$, which determine the selection
rule for the optical response. A tedious manipulation leads us to
\begin{equation}
{\cal I}_{m\bar{m}}=\lambda_{J_{z}}\lambda_{M}{\cal I}_{m\bar{m}},
\end{equation}
where $\lambda_{{\cal O}}$ is the parity of the operator ${\cal O}$
with respect to ${\cal A}$. Thus, if $\lambda_{J_{z}}\lambda_{M}=-1$,
the linear coupling between the pairing fluctuation and the light
characterized by the form factor $M_{\boldsymbol{k}}$ is forbidden.
In Table \ref{tab:CharacterTable}, the parity of the form factors
of the pairing channels are listed. Note that both $J_{z}$ and $\sigma_{z}$
are odd under ${\cal A}$ while $k_{x}s_{y}-k_{y}s_{x}$ is even.
Therefore, the linear coupling between the light and the fluctuation
in the pairing channel $k_{x}s_{y}-k_{y}s_{x}$ is negligible as long
as the Ising-type spin-orbit coupling is negligible.

\subsection{Both superconducting phases are odd-parity under inversion}

In Ref. \citep{Mockli2021}, it is proposed that the $H-T$ phase
diagram of the superconducting states of ${\rm CeRh_{2}As_{2}}$ might
be reproduced with inter-layer spin-triplet odd-parity gap functions.
There, the low-field state is characterized by an odd-parity spin-triplet
gap function transforming like $k_{x}k_{y}k_{z}(k_{x}^{2}-k_{y}^{2})\sigma_{x}s_{z}$
that belongs to $A_{1u}$ of $D_{4h}$. The gap function of the high-field
state is another odd-parity spin-triplet gap function transforming
$\sigma_{y}s_{z}$ belonging to $A_{2u}$ of $D_{4h}$.

For this case, a BS mode should exist because both pairing channels
belong to different irreducible representation. Since both pairing
channels have the same inversion parity, the BS mode is inactive in
the linear optical response.

\section{Odd-parity superconductivity in the in-plane magnetic fields \label{App:oSC-Bx}}

The behavior of the gap functions $\Delta_{e}$ and $\Delta_{o}$
under the in-plane magnetic field is exposed in detail here.

\subsection{eSC phase under the in-plane magnetic field $B_{x}$}

Using the basis diagonalizing $H_{0}(\boldsymbol{k})$, the BdG Hamiltonian
in $p$SC is written as
\begin{align}
U^{\dagger}H_{\text{BdG}}^{(e)}U= & \tau_{z}\left(\begin{matrix}\xi_{1}\\
 & \xi_{2}
\end{matrix}\right)+\Delta_{e}\tau_{x}+B_{x}U^{\dagger}s_{x}U,\label{Appeq:UHeU}\\
U^{\dagger}s_{x}U= & \left(\begin{matrix}\hat{\boldsymbol{a}}\cdot\boldsymbol{\rho} & -i\rho_{0}\cos\chi\sin\phi\\
i\rho_{0}\cos\chi\sin\phi & -\hat{\boldsymbol{a}}\cdot\boldsymbol{\rho}
\end{matrix}\right),
\end{align}
where 
\begin{equation}
U=\frac{1}{\sqrt{2}}\tau_{0}\otimes\left(\begin{matrix}1 &  & 1\\
a & a^{*}e^{i\zeta} & -b & -b^{*}e^{i\zeta}\\
b & -b^{*}e^{i\zeta} & -a & a^{*}e^{i\zeta}\\
 & e^{i\zeta} &  & e^{i\zeta}
\end{matrix}\right).\label{Appeq:U}
\end{equation}
with $a=e^{i\phi}\sin\chi$ and $b=e^{i\zeta}\cos\chi$. Here, $\rho_{i}$
are the pseudospin Pauli matrices and $\hat{\boldsymbol{a}}=(\cos\chi,0,\sin\chi\cos\phi)$.
Assuming that $|B_{x}|\ll|\xi_{1}-\xi_{2}|$ at both Fermi surfaces,
we can ignore the off-diagonal components of $U^{\dagger}s_{x}U$
in the rhs of Eq. \eqref{Appeq:UHeU}. To gain a meaningful insight,
we further introduce another unitary matrix $U_{a}$ that diagonalizes
$\hat{\boldsymbol{a}}\cdot\boldsymbol{\rho}$. Then, the BdG Hamiltonian
$U_{a}^{\dagger}U^{\dagger}H_{\text{BdG}}^{(e)}UU_{a}$ is approximated
by $\bar{H}_{\text{BdG}}^{(e)}$ given by
\begin{align}
\bar{H}_{\text{BdG}}^{(e)}= & \tau_{z}\bigg(\begin{matrix}\xi_{1}\\
 & \xi_{2}
\end{matrix}\bigg)+\mathfrak{g}B_{x}\tau_{0}\bigg(\begin{matrix}\mu_{z}\\
 & -\mu_{z}
\end{matrix}\bigg)+\tau_{x}\Delta_{e}.\label{Appeq:barHe}
\end{align}
Here, $\mu_{i}$ are the Pauli matrices the final basis and $\mathfrak{g}=\sqrt{\cos^{2}\phi\sin^{2}\chi+\cos^{2}\chi}$
represents an\emph{ }effective\emph{ }Zeeman coupling in response
to the in-plane magnetic field.

Note that $\bar{H}_{\text{BdG}}^{(e)}$ is decomposed into four $2\times2$
blocks which take the following form $\pm\tau_{0}\mathfrak{g}B_{x}+\tau_{z}\xi_{i}+\tau_{x}\Delta_{e}$
whose eigenvalues are given by $\pm B_{x}\mathfrak{g}\pm\sqrt{\xi_{i}^{2}+\Delta_{e}^{2}}$.
Direct manipulation of the self-consistent gap equation results in
\begin{align}
\frac{1}{g_{e}} & =\sum_{\boldsymbol{k}}\sum_{i}\frac{1}{2E_{i}}\sum_{s=\pm}\tanh\frac{\beta(E_{i}+s\mathfrak{g}B_{x})}{2},
\end{align}
which is reduced to the following form at the zero-temperature
\begin{equation}
\frac{1}{g_{e}}=\sum_{\boldsymbol{k}}\sum_{i}\frac{\Theta(E_{i}-\mathfrak{g}B_{x})}{E_{i}}.\label{Appeq:SCE_in_eSC}
\end{equation}
The Heaviside theta function appear because of the cancellation of
the two $\tanh$'s in Eq. \eqref{Appeq:SCE_in_eSC}. Indeed, this
equation is exactly what explains the first-order transition by the
Pauli pair-breaking in the conventional superconductors \citep{Sarma1963}.
However, if $B_{x}$ is increased so that $B_{x}\text{min}_{\phi}(\mathfrak{g})>\Delta_{e}$,
then a low-energy region $|\xi_{i}|<\sqrt{B_{x}^{2}\min_{\phi}(\mathfrak{g}^{2})-\Delta_{e}^{2}}$
is got rids of from the energy integration, which prevents us to obtain
a finite $\Delta_{e}$ as a solution of Eq. \eqref{Appeq:SCE_in_eSC}
for small $g_{e}$. Figure. \ref{Appfig:App}(a) shows $\Delta_{e}$
which is obtained by numerically solving the self-consistent gap equation
under the in-plane magnetic fields. It is easily recognized that $\Delta_{e}$
discretely jumps to zero for strong $B_{x}$.

\subsection{oSC phase under the in-plane magnetic field $B_{x}$}

Using $U$ in Eq. \eqref{Appeq:U}, the oSC BdG Hamiltonian is transformed
into
\begin{align}
U^{\dagger}H_{\text{BdG}}^{(o)}U= & \tau_{z}\bigg(\begin{matrix}\xi_{1}\\
 & \xi_{2}
\end{matrix}\bigg)+U^{\dagger}\{\Delta_{o}\tau_{x}\sigma_{z}+B_{x}s_{x}\}U,
\end{align}
where 
\begin{equation}
U^{\dagger}\sigma_{z}U=\left(\begin{matrix}\sin\chi\hat{\boldsymbol{\Sigma}}_{d}\cdot\boldsymbol{\rho} & \cos\chi\hat{\boldsymbol{\Sigma}}_{od}\cdot\boldsymbol{\rho}\\
\cos\chi\hat{\boldsymbol{\Sigma}}_{od}\cdot\boldsymbol{\rho} & \sin\chi\hat{\boldsymbol{\Sigma}}_{d}\cdot\boldsymbol{\rho}
\end{matrix}\right)
\end{equation}
with $\hat{\boldsymbol{\Sigma}}_{d}=(\cos\chi\cos\phi,\cos\chi\sin\phi,\sin\chi)$
and $\hat{\boldsymbol{\Sigma}}_{od}=(-\sin\chi\sin\phi,-\sin\chi\sin\phi,\cos\chi)$.
Adopting the same assumptions used in analyzing the eSC phase, we
neglect the off-diagonal blocks in $U^{\dagger}\sigma_{z}U$ and $U^{\dagger}s_{x}U$.
We also neglect the off-diagonal components in the diagonal blocks
of $B_{x}U^{\dagger}s_{x}U$. The final BdG Hamiltonian that we use
to address the behavior of $\Delta_{o}$ under the in-plane magnetic
field is given by \begin{widetext}
\begin{align}
U^{\dagger}H_{\text{BdG}}^{(o)}U\approx\bar{H}_{\text{BdG}}^{(o)}= & \left(\begin{matrix}\xi_{1}+\bar{B}_{x}\rho_{z} &  & \Delta_{o}\sin\chi\hat{\boldsymbol{\Sigma}}_{d}\cdot\boldsymbol{\rho}\\
 & \xi_{2}-\bar{B}_{x}\rho_{z} &  & \Delta_{o}\sin\chi\hat{\boldsymbol{\Sigma}}_{d}\cdot\boldsymbol{\rho}\\
\Delta_{o}\sin\chi\hat{\boldsymbol{\Sigma}}_{d}\cdot\boldsymbol{\rho} &  & -\xi_{1}+\bar{B}_{x}\rho_{z}\\
 & \Delta_{o}\sin\chi\hat{\boldsymbol{\Sigma}}_{d}\cdot\boldsymbol{\rho} &  & -\xi_{2}-\bar{B}_{x}\rho_{z}
\end{matrix}\right)\\
= & \left(\begin{matrix}\xi_{1}+\bar{B}_{x}\rho_{z} & \Delta_{o}\sin\chi\hat{\boldsymbol{\Sigma}}_{d}\cdot\boldsymbol{\rho}\\
\Delta_{o}\sin\chi\hat{\boldsymbol{\Sigma}}_{d}\cdot\boldsymbol{\rho} & -\xi_{1}+\bar{B}_{x}\rho_{z}
\end{matrix}\right)\oplus\left(\begin{matrix}\xi_{2}+\bar{B}_{x}\rho_{z} & \Delta_{o}\sin\chi\hat{\boldsymbol{\Sigma}}_{d}\cdot\boldsymbol{\rho}\\
\Delta_{o}\sin\chi\hat{\boldsymbol{\Sigma}}_{d}\cdot\boldsymbol{\rho} & -\xi_{2}+\bar{B}_{x}\rho_{z}
\end{matrix}\right)\label{Appeq:barHo}
\end{align}
\end{widetext}with $\bar{B}_{x}=B_{x}\cos\phi\sin\chi$.

It is noteworthy that $\bar{H}_{\text{BdG}}^{(o)}$, contrasting to
$\bar{H}_{\text{BdG}}^{(e)}$, involves a pairing between two bands
of the same energy dispersion $\xi_{i}\pm\sin\chi\cos\phi B_{x}$
through the off-diagonal component in $\Delta_{o}\sin\chi\hat{\boldsymbol{\Sigma}}_{d}\cdot\boldsymbol{\rho}$.
If it were not for the off-diagonal elements, the overall structure
of $\bar{H}_{\text{BdG}}^{(o)}$ would be identical to that of $\bar{H}_{\text{BdG}}^{(e)}$
in Eq. \eqref{Appeq:barHe}, and the oSC phase would exhibit a discontinuous
transition to the normal phase at a strong enough $B_{x}$. The off-diagonal
components in $\Delta_{o}\sin\chi\hat{\boldsymbol{\Sigma}}_{d}\cdot\boldsymbol{\rho}$
make the difference. Let us investigate their implication by neglecting
the diagonal components completely. Under this assumption, the $4\times4$
subblocks of $\bar{H}_{\text{BdG}}^{(o)}$ is further decomposed into
two $2\times2$ subblocks of the form like
\begin{equation}
\left(\begin{matrix}\xi_{i}+\sin\chi\cos\phi B_{x} & \Delta_{o}\cos\chi\sin\chi e^{i\phi}\\
\Delta_{o}\cos\chi\sin\chi e^{-i\phi} & -\xi_{i}-\sin\chi\cos\phi B_{x}
\end{matrix}\right).
\end{equation}
Note that $B_{x}\sin\chi\cos\phi$ can be absorbed into $\xi_{i}$
and thus the fully gapped superconductivity is retained for any $g_{o}$
regardless of $B_{x}$.

\begin{figure}
\includegraphics[width=1\columnwidth]{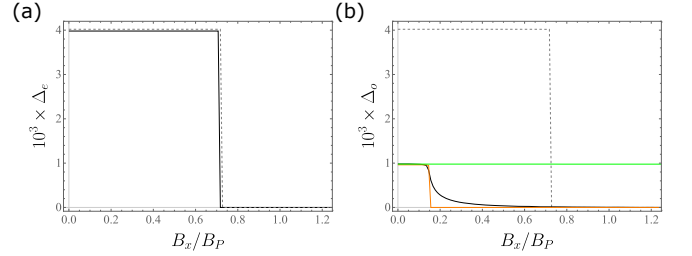}

\caption{\label{Appfig:App}The evolution of the gap functions under the increasing
in-plane magnetic fields. (a) The solid (dashed) line represent $\Delta_{e}$
vs $B_{x}$ when $t_{c}=0.1$ ($t_{c}=0$) is used in $H_{0}$. The
$t_{c}=0$ case corresponds to the purely 2D case and largely coincides
with the result in Ref. \citep{Tewari2011}. (b) $\Delta_{o}$ vs
$B_{x}$ for several cases. The gray dashed line is obtained with
$t_{c}=0.0005\ll|\Delta_{o}|$ in $H_{0}$, which is qualitatively
same with the dashed line in (a). The solid lines are obtained with
$t_{c}=0.1\gg|\Delta_{o}|$. Of the three lines, the block line corresponds
to a solution of the self-consistent gap equation when every terms
in $H_{\text{BdG}}^{(o)}$ are retained. The green and orange lines
are the results calculated by using $\bar{H}_{\text{BdG}}^{(o)}$
in Eq. \eqref{Appeq:barHo} neglecting the diagonal and the off-diagonal
components in the gap function part, respectively.}
\end{figure}

The numerical results displayed in Fig. \ref{Appfig:App}(b) confirm
the analytical analysis. The green and orange lines represents solutions
of the self-consistent gap equation, which are obtained by using $\bar{H}_{\text{BdG}}^{(o)}$
without the diagonal and the off-diagonal components in the gap functions,
respectively. With only the diagonal components in the gap function,
a first-order transition to the normal phase appears, while the superconducting
phase can robustly withstand against the in-plane magnetic fields
when only the off-diagonal components are retained. The black solid
line represents a solution of the self-consistent gap equation using
$H_{\text{BdG}}^{(o)}$ without any approximation. The solution shows
an intriguing exponential decrease under $B_{x}$, which is a compromise
between what the diagonal and off-diagonal components favor. Because
of the exponential decrease which never touches the zero, the Pauli
limiting field $B_{x,\text{P}}$ in the oSC state is infinite at the
zero temperature, which is in a sharp contrast to the eSC state. The
gray dashed line is a solution of the self-consistent gap equation
with $H_{\text{BdG}}^{(o)}$ when the interlayer hopping $t_{c}$
in $H_{0}$ is set to a value much smaller than the gap function.
In this 2D limit, the sign-alternating gap function $\sigma_{z}$
is not discerned from the trivial gap function $\sigma_{0}$ by the
electrons. Thus, the oSC state also exhibits the first-order transition
to the normal phase like as the eSC state does.

\section{Intraband components of $\langle m|{\cal V}_{i}|n\rangle$ in the
trivial superconducting state\label{App:Intraband-components}}

In this section, we are going to show explicitly that the components
of $\langle m|{\cal V}_{i}|n\rangle$ corresponding to the intraband
transitions in the normal phase band structure are zero following
Ref. \citep{Ahn2021}. Using the eigenvectors in Eq. \eqref{eq:CeRhAs_BdG_Eigenvector}
of the BdG Hamiltonian $H_{\text{BdG}}^{(e)}$, the components of
$\langle m|{\cal V}_{i}|n\rangle$ which are relevant to the calculation
of $L_{z,2}^{(p)}$ at the zero temperature are given by
\begin{equation}
\langle c,i,\alpha|\partial_{l}H|v,j,\alpha'\rangle=\sin\frac{\Xi_{i}-\Xi_{j}}{2}\langle i,\alpha|\partial_{l}H_{0}|j,\alpha'\rangle.
\end{equation}
Note that the right-hand side is zero whenever $\Xi_{i}=\Xi_{j}$,
which is satisfied if $\xi_{i}=\xi_{j}$. Therefore, no components
$\langle m|{\cal V}_{i}|n\rangle$ corresponding to the intraband
transitions in the normal phase band structure are finite. The proof
can be generalized to the case in which the gap function commutes
with the normal phase Hamiltonian and preserves the spatio-temporal
inversion so that the eigenstates degenerate due to the Kramers' theorem
are subject to the gap functions of the same magnitude.

\section{Estimation of $L_{z,2}^{(o)}$ when $g_{e}=g_{o}$\label{App:Estimation_Lz2}}

When $g_{o}=g_{e}$, we can use $\langle\sin^{2}\chi\rangle_{\text{FS}}=\frac{\ln(\Lambda/T_{c,e})}{\ln(\Lambda/T_{c,o})}$,
and it is possible to find a simple relation between $L_{z,2}^{(o)}(\Omega)$
and the superconducting transition temperatures $T_{c,e}$ and $T_{c,o}$.
Using the expressions for $T_{c,e}$ and $T_{c,o}$ in the weak-coupling
limit, $L_{z,2}^{(o)}(\Omega)$ is expressed by
\begin{align}
L_{z,2}^{(o)}(\Omega)\approx & -\frac{1.73N_{\text{tot}}T_{c,e}}{2}\frac{\ln(\Lambda/T_{c,e})}{\ln(\Lambda/T_{c,o})}\ln\frac{T_{c,e}}{T_{c,o}}\nonumber \\
\approx & -\frac{1.73N_{\text{tot}}T_{c,e}}{2}\ln\frac{T_{c,e}}{T_{c,o}},\label{eq:Loze_with_Tc}
\end{align}
where $\Delta_{e}=1.73T_{c,e}$ and $\langle\sin^{2}\chi\rangle_{\text{FS}}=\frac{\ln(\Lambda/T_{c,e})}{\ln(\Lambda/T_{c,o})}$
are used to obtain the first line of Eq. \eqref{eq:Loze_with_Tc}
and $\Lambda\gg T_{c,e},T_{c,o}$ is used to obtain the second line.

\section{Analytical expressions for $\Pi^{(o)}(\Omega)$ in the eSC state
and $\Omega_{\text{BS}}$\label{App:Pi_Omega_Bash-even}}

In this section, an analytical expression for the gap of the BS mode
from a subdominant odd parity pairing channel is presented. In general,
the gap of this collective mode can be determined by solving $\det[g_{o}^{-1}+\Pi^{(o)}(\Omega)]=0$
within the generalized random-phase approximation formalism, and we
are going to calculate each elements of $\Pi^{(o)}(\Omega)$ for $|\Omega|<2\Delta_{e}$
in appropriate approximations. To get to the point first,
\begin{align}
\Pi_{11}^{(o)}(\Omega) & =\langle\sin^{2}\chi\rangle\bigg\{\frac{2N_{\text{tot}}\arcsin y}{y/\sqrt{1-y^{2}}}-\frac{1}{g_{e}}\bigg\},\\
\Pi_{12}^{(o)}(\Omega) & \approx\langle\cos^{2}\chi\rangle\Pi_{12}^{(e)}(\Omega),\\
\Pi_{22}^{(o)}(\Omega) & \approx-\langle\sin^{2}\chi\rangle\bigg\{\frac{1}{g_{e}}+N_{\text{tot}}\frac{2y\arcsin y}{\sqrt{1-y^{2}}}\bigg\}.
\end{align}
with $y=\Omega^{+}/2\Delta_{e}$ and $N_{\text{tot}}=N_{1}+N_{2}$,
which is used throughout this section. Here, $N_{i}$ is the density
of states at the Fermi surface from the band $\xi_{i}$. (Counting
the Kramer degeneracy, $2N_{\text{tot}}$ is the total density of
states.)

The following formulae are frequently used in the derivation.
\begin{align}
|\langle\xi_{i},s|\sigma_{z}|\xi_{j},s'\rangle| & =(1-\delta_{ss'})\begin{cases}
|\sin\chi| & i=j\\
|\cos\chi| & i\neq j
\end{cases},\\
\langle c,i,s|\tau_{y}^{(o)}|v,j,s'\rangle & =\frac{1}{i}\cos\frac{\Xi_{j}-\Xi_{i}}{2}\langle\xi_{i},s|\sigma_{z}|\xi_{j},s'\rangle,\\
\langle c,i,s|\tau_{x}^{(o)}|v,j,s'\rangle & =\cos\frac{\Xi_{j}+\Xi_{i}}{2}\langle\xi_{i},s|\sigma_{z}|\xi_{j},s'\rangle,\\
\frac{1}{g_{e}} & =\sum_{i}N_{i}\int_{-\Lambda}^{\Lambda}\text{d}\xi_{i}\frac{1}{E_{c,i}},
\end{align}
for $i,j=1,2$ and $s,s'=\alpha,\beta$ derived by using the eigenvectors
of $H_{\text{BdG}}$ in Eq. \eqref{eq:CeRhAs_BdG_Eigenvector}. The
last equation is the BCS gap equation with the energy cut-off $\Lambda$.

\subsection{$\Pi_{22}^{(o)}(\Omega)$ in the eSC state}

For later use, we first evaluate $\Pi_{22}^{(o)}(0)$ using the identity
\begin{align}
[\tau_{z}\sigma_{z},G_{k}^{-1}] & =-2i\tau_{y}^{(o)}\Delta_{e,\text{R}}-4iF_{A}.
\end{align}
where $4iF_{A}\equiv[\sigma_{z},H_{0}]$. This commutator appears
in literature which are concerned with the concept of superconducting
fitness \citep{Mockli2021}. When the Ising-type spin-orbit coupling
is ignored ($\lambda_{\text{Ising}}=0$) in Eq. \eqref{eq:CeRhAs-normal},
$F_{A}=J_{z}$ is satisfied. Given the self-consistent gap equation
$2\Delta_{e,\text{R}}=-g_{e}\text{Tr}[\tau_{x}G_{k}]$ and $F_{A}=J_{z}$,
we get 
\begin{align}
\Pi_{22}^{(o)}(0)= & -\frac{1}{g_{e}}-\sum_{k}\frac{\text{Tr}[\tau_{y}\sigma_{z}G(k)F_{A}(\boldsymbol{k})G(k)]}{\Delta_{e,\text{R}}}\\
= & -\frac{1}{g_{e}}-\frac{2L_{z,2}^{(o)}(0)}{\Delta_{e,\text{R}}}\approx-\frac{\langle\sin^{2}\chi\rangle}{g_{e}},
\end{align}
where we use Eq. \eqref{eq:Loze} for the last approximation.

Now, we calculate $\Pi_{22}^{(o)}(\Omega)$. First, $\Pi_{22}^{(o)}(\Omega)$
at zero-temperature is written as
\begin{align}
\Pi_{22}^{(o)}(\Omega)= & \Pi_{22}^{(o,\text{sig})}(\Omega)+\Pi_{22}^{(o,\text{reg})}(\Omega),\label{eq:Pi_o_22}
\end{align}
with
\begin{align}
\Pi_{22}^{(o,\text{sig})}(\Omega)= & \check{\sum_{\boldsymbol{k}}}\sum_{i=1,2}\frac{4E_{c,i}\sin^{2}\chi}{(\Omega^{+})^{2}-4E_{c,i}^{2}},\\
\Pi_{22}^{(o,\text{reg})}(\Omega)= & \check{\sum_{\boldsymbol{k}}}\frac{4\cos^{2}\chi\cos^{2}\frac{\Xi_{1}-\Xi_{2}}{2}(E_{c,1}+E_{c,2})}{(\Omega^{+})^{2}-(E_{c,1}+E_{c,2})^{2}},\nonumber 
\end{align}

The potential singularities of $\Pi_{22}^{(o,\text{reg})}(\Omega)$
lie at $\Omega=E_{c,1}+E_{c,2}\gg2\Delta$, and thus $\Pi_{22}^{(o,\text{reg})}(\Omega)\approx\Pi_{22}^{(o,\text{reg})}(0)$
is a good approximation for $|\Omega|<2\Delta_{e}$. A tedious manipulation
easily leads to 
\begin{equation}
\Pi_{22}^{(o,\text{reg})}(0)+\check{\sum_{\boldsymbol{k},i}}\frac{\cos^{2}\chi}{E_{c,i}}=\frac{2L_{z,2}^{(o)}(0)}{-\Delta_{e}}=\frac{\langle\cos^{2}\chi\rangle}{g_{e}},
\end{equation}
implying $\Pi_{22}^{(o,\text{reg})}(0)\approx0$. In this approximation,
$\Pi_{22}^{(o)}(\Omega)-\Pi_{22}^{(o)}(0)$ turns out to be 
\begin{align}
\Pi_{22}^{(o)}(\Omega)-\Pi_{22}^{(o)}(0)= & \check{\sum_{\boldsymbol{k}}}\sum_{i}\frac{(\Omega^{+})^{2}\sin^{2}\chi}{E_{c,i}((\Omega^{+})^{2}-4E_{c,i}^{2})}\nonumber \\
= & -\langle\sin^{2}\chi\rangle N_{\text{tot}}\frac{2y\arcsin y}{\sqrt{1-y^{2}}}.
\end{align}
Therefore, we obtain
\begin{equation}
\Pi_{22}^{(o)}(\Omega)=-\langle\sin^{2}\chi\rangle\bigg(\frac{1}{g_{e}}+\frac{2N_{\text{tot}}y\arcsin y}{\sqrt{1-y^{2}}}\bigg).
\end{equation}
Note that this function diverges at $y=1$.

\subsection{$\Pi_{12}^{(o)}(\Omega)$ in the eSC state}

$\Pi_{12}^{(o)}(\Omega)$ at the zero-temperature is written as
\begin{align}
\Pi_{12}^{(o)}(\Omega)= & \Pi_{12}^{(o,\text{sing})}(\Omega)+\Pi_{12}^{(o,\text{reg})}(\Omega),\label{eq:Pi_o_12}
\end{align}
with 
\begin{align}
\Pi_{12}^{(o,\text{sig})}(\Omega)= & \check{\sum_{\boldsymbol{k}}}\sum_{i=1,2}\frac{2i\Omega^{+}\cos\Xi_{i}\sin^{2}\chi}{4E_{c,i}^{2}-(\Omega^{+})^{2}},\\
\Pi_{12}^{(o,\text{reg})}(\Omega)= & \check{\sum_{\boldsymbol{k}}}\frac{2i\Omega^{+}(\cos\Xi_{1}+\cos\Xi_{2})\cos^{2}\chi}{(E_{c,1}+E_{c,2})^{2}-(\Omega^{+})^{2}}.\nonumber 
\end{align}
Note that $\Pi_{12}^{(o,\text{sig})}=\langle\sin^{2}\chi\rangle\Pi_{12}^{(e)}(\Omega)$
which is is a linear function of $\Omega$ and only finite when the
electronic density of states is asymmetric around $\xi_{i}=0$. Meanwhile,
$\Pi_{12}^{(o,\text{reg})}(\Omega)$ is negligible in the weak-coupling
theory because $\cos\Xi_{1}+\cos\Xi_{2}\sim1$ and $(E_{c,1}+E_{c,2})^{2}-(\Omega^{+})^{2}\sim4(\xi_{1}-\xi_{2})^{2}\gg\Lambda\Delta$
with $\Lambda$ the integration cutoff. Thus, we neglect it, and obtain
\begin{align}
\Pi_{12}^{(o)}(\Omega) & \approx\langle\sin^{2}\chi\rangle\Pi_{12}^{(e)}(\Omega).
\end{align}

\subsection{$\Pi_{11}^{(o)}(\Omega)$ in the eSC state}

$\Pi_{11}^{(e)}(\Omega)$ at the zero-temperature is written as
\begin{align}
\Pi_{11}^{(o)}(\Omega)= & \Pi_{11}^{(o,\text{sig})}(\Omega)+\Pi_{11}^{(o,\text{reg})}(\Omega),\label{eq:Pi_o_11}
\end{align}
with 
\begin{align}
\Pi_{11}^{(o,\text{sig})}(\Omega)= & \check{\sum_{\boldsymbol{k}}}\sum_{i=1,2}\frac{4\xi_{i}^{2}\sin^{2}\chi}{E_{c,i}\{(\Omega^{+})^{2}-4E_{c,i}^{2}\}},\\
\Pi_{11}^{(o,\text{reg})}(\Omega)= & \check{\sum_{\boldsymbol{k}}}\frac{2(E_{c,1}+E_{c,2})\cos^{2}\chi}{(\Omega^{+})^{2}-(E_{c,1}+E_{c,2})^{2}}Z,
\end{align}
with $Z=(E_{1}E_{2}+\xi_{1}\xi_{2}-\Delta^{2})/(2E_{1}E_{2})$. $\Pi_{11}^{(o,\text{reg})}(\Omega)$
is negligible since it is order of $(\Delta/(\xi_{1}-\xi_{2}))^{2}$.
Evaluating the singular term $\Pi_{11}^{(o,\text{sig})}(\Omega)$
at $\Omega=0$, we get
\begin{align}
\Pi_{11}^{(o)}(0) & =-\frac{\langle\sin^{2}\chi\rangle}{g_{e}}+2\langle\sin^{2}\chi\rangle N_{\text{tot}}.
\end{align}
Calculating $\Pi_{11}^{(o)}(\Omega)-\Pi_{11}^{(o)}(0)$, we obtain
\begin{align}
\Pi_{11}^{(o)}(\Omega)-\Pi_{11}^{(o)}(0)= & \check{\sum_{\boldsymbol{k}}}\sum_{i}\frac{\xi_{i}^{2}\sin^{2}\chi(\Omega^{+})^{2}}{E_{c,i}^{3}\{(\Omega^{+})^{2}-4E_{c,i}^{2}\}}\\
= & 2\langle\sin^{2}\chi\rangle N_{\text{tot}}\bigg(\frac{\arcsin y}{y/\sqrt{1-y^{2}}}-1\bigg),\nonumber 
\end{align}
which leads to 
\begin{equation}
\Pi_{11}^{(o)}(\Omega)=-\frac{\langle\sin^{2}\chi\rangle}{g_{e}}+\langle\sin^{2}\chi\rangle N_{\text{t ot}}\frac{2\arcsin y}{y/\sqrt{1-y^{2}}}.
\end{equation}
Note that $\Pi_{11}^{(o)}(0)>\Pi_{11}^{(o)}(2\Delta)=\Pi_{22}^{(o)}(0)$
\citep{Sauls2017,Nambu1985}, which implies $\Pi_{11}^{(o)}(\Omega)+\frac{1}{g_{o}}$
is a regular function for $|\Omega|<2\Delta$ because $\Pi_{11}^{(o)}(\Omega)+\frac{1}{g_{o}}>0$
when $\Pi_{22}^{(o)}(\Omega)+\frac{1}{g_{o}}=0$.

\subsection{$\Omega_{\text{BS}}$ in the eSC state}

The gap of the BS mode is obtained by finding the zero of $\det[g_{o}^{-1}+\Pi^{(0)}(\Omega)]=0$.
Since $\Pi_{11}^{(o)}(\Omega)$ is regular and finite for $|\Omega|<2\Delta$
and $\Pi_{12}^{(o)}(\Omega)$ and $\Pi_{21}^{(o)}$ are expected to
be small as we have shown, $\Omega_{\text{BS}}$ can be accurately
approximated by the zero of $\Pi_{22}^{(0)}(\Omega)+g_{o}^{-1}=0$.Consequently,
the semi-analytical expression for $\Omega_{\text{BS}}$ is given
by 
\begin{equation}
\frac{2y\arcsin y}{\sqrt{1-y^{2}}}=\frac{1}{\langle\sin^{2}\chi\rangle g_{o}N_{\text{tot}}}-\frac{1}{g_{e}N_{\text{tot}}}=\log\frac{T_{c,e}}{T_{c,o}},\label{eq:BS-mode}
\end{equation}
where the last equality is derived under the assumption that the superconducting
transition temperatures $T_{c,e}$ and $T_{c,o}$ take the BCS-like
form $T_{c,e}\sim\Lambda\exp[-1/g_{e}N_{\text{tot}}]$ and $T_{c,o}\sim\Lambda\exp[-1/g_{e}\langle\sin^{2}\chi\rangle N_{\text{tot}}]$,
respectively.

For $T_{c,o}=0.87T_{c,e}$, Eq. \eqref{eq:BS-mode} yields $y=\Omega_{\text{BS}}/2\Delta_{e}\sim0.26$
in the eSC state. Therefore, the BS mode in the eSC state is expected
to be found below the midst of the superconducting excitation gap.

\subsection{The peak intensity from the BS mode in $\tilde{K}_{zz}$}

Here, we evaluate the intensity of the Dirac delta peak in $\tilde{K}_{zz}$
from the BS mode. The peak is mainly attributed to the zero of $\Pi_{22}^{(o)}(\Omega)+g_{o}^{-1}=0$,
and thus we approximate $[L^{(o)}(g_{o}^{-1}+\Pi^{(o)})^{-1}R^{(o)}]_{z,z}$
by 
\begin{equation}
[L^{(o)}(g_{o}^{-1}+\Pi^{(o)})^{-1}R^{(o)}]_{z,z}\approx\frac{L_{z,2}^{(o)}R_{2,z}^{(o)}}{\frac{1}{g_{o}}+\Pi_{22}^{(o)}(\Omega)+i\epsilon}.
\end{equation}
As the imaginary part of $\Pi_{22}^{(o)}(\Omega)$ should be positive
for causality, we add an infinitesimal number $\epsilon$ to $\Pi_{22}^{(o)}(\Omega)$.
Taking the limit $\epsilon\rightarrow0^{+}$, we get the Dirac delta
peak of the BS mode
\begin{align}
\frac{L_{z,2}^{(o)}R_{2,z}^{(o)}}{\frac{1}{g_{o}}+\Pi_{22}^{(o)}(\Omega)+i\epsilon}= & |L_{z,2}^{(o)}|^{2}\frac{\delta(\Omega-\Omega_{\text{BS}})}{\partial_{\Omega}\Pi_{22}^{(o)}\big|_{\Omega_{\text{BS}}}}\\
= & \frac{\Omega_{\text{BS}}|L_{z,2}^{(o)}|^{2}}{2\langle\sin^{2}\chi\rangle N_{\text{tot}}}p(\Omega_{\text{BS}})\delta(\Omega-\Omega_{\text{BS}}),\nonumber 
\end{align}
with $|L_{z,2}^{(o)}|^{2}=\frac{\Delta^{2}\langle\cos^{2}\chi\rangle^{2}}{4g_{e}^{2}}$
and 
\begin{equation}
p(\Omega_{\text{BS}})=\frac{1-y_{\text{BS}}^{2}}{\frac{1}{2N_{\text{tot}}}\bigg\{\frac{1}{g_{o}\langle\sin^{2}\chi\rangle}-\frac{1}{g_{e}}\bigg\}+y_{\text{BS}}^{2}}.
\end{equation}
If the solution $y_{\text{BS}}$ of Eq. \eqref{eq:BS-mode} is small,
it can be approximated by $2g_{e}N_{\text{tot}}y_{\text{BS}}^{2}\approx\frac{g_{e}}{g_{o}\langle\sin^{2}\chi\rangle}-1$.
Thus, $p(\Omega_{\text{BS}})$ is also expected to scale like $y_{\text{BS}}^{-2}$
for small $y_{\text{BS}}$ and the intensity is fortified like $1/\Omega_{\text{BS}}$
for small $\Omega_{\text{BS}}$.

This scaling of the peak intensity for small $\Omega_{\text{BS}}$
is a consequence of the generic properties of the linear response
functions rendering $\det[1/g_{o}^{-1}+\Pi^{(o)}(\Omega)]$ an even
function of $\Omega$ for $|\Omega|<2\Delta_{e}$. Therefore, regardless
of the control parameter used to lower $\Omega_{\text{BS}}$, the
peak intensity, or height, will increase as $\Omega_{\text{BS}}$
is lowered.

\bibliography{ref}

\begin{thebibliography}{46}%
\makeatletter
\providecommand \@ifxundefined [1]{%
 \@ifx{#1\undefined}
}%
\providecommand \@ifnum [1]{%
 \ifnum #1\expandafter \@firstoftwo
 \else \expandafter \@secondoftwo
 \fi
}%
\providecommand \@ifx [1]{%
 \ifx #1\expandafter \@firstoftwo
 \else \expandafter \@secondoftwo
 \fi
}%
\providecommand \natexlab [1]{#1}%
\providecommand \enquote  [1]{``#1''}%
\providecommand \bibnamefont  [1]{#1}%
\providecommand \bibfnamefont [1]{#1}%
\providecommand \citenamefont [1]{#1}%
\providecommand \href@noop [0]{\@secondoftwo}%
\providecommand \href [0]{\begingroup \@sanitize@url \@href}%
\providecommand \@href[1]{\@@startlink{#1}\@@href}%
\providecommand \@@href[1]{\endgroup#1\@@endlink}%
\providecommand \@sanitize@url [0]{\catcode `\\12\catcode `\$12\catcode
  `\&12\catcode `\#12\catcode `\^12\catcode `\_12\catcode `\%12\relax}%
\providecommand \@@startlink[1]{}%
\providecommand \@@endlink[0]{}%
\providecommand \url  [0]{\begingroup\@sanitize@url \@url }%
\providecommand \@url [1]{\endgroup\@href {#1}{\urlprefix }}%
\providecommand \urlprefix  [0]{URL }%
\providecommand \Eprint [0]{\href }%
\providecommand \doibase [0]{https://doi.org/}%
\providecommand \selectlanguage [0]{\@gobble}%
\providecommand \bibinfo  [0]{\@secondoftwo}%
\providecommand \bibfield  [0]{\@secondoftwo}%
\providecommand \translation [1]{[#1]}%
\providecommand \BibitemOpen [0]{}%
\providecommand \bibitemStop [0]{}%
\providecommand \bibitemNoStop [0]{.\EOS\space}%
\providecommand \EOS [0]{\spacefactor3000\relax}%
\providecommand \BibitemShut  [1]{\csname bibitem#1\endcsname}%
\let\auto@bib@innerbib\@empty
\bibitem [{\citenamefont {Joynt}\ and\ \citenamefont
  {Taillefer}(2002)}]{Joynt2002}%
  \BibitemOpen
  \bibfield  {author} {\bibinfo {author} {\bibfnamefont {R.}~\bibnamefont
  {Joynt}}\ and\ \bibinfo {author} {\bibfnamefont {L.}~\bibnamefont
  {Taillefer}},\ }\bibfield  {title} {\bibinfo {title} {{The superconducting
  phases of $\rm{U Pt_3}$}},\ }\href
  {https://doi.org/10.1103/RevModPhys.74.235} {\bibfield  {journal} {\bibinfo
  {journal} {Rev. Mod. Phys.}\ }\textbf {\bibinfo {volume} {74}},\ \bibinfo
  {pages} {235} (\bibinfo {year} {2002})}\BibitemShut {NoStop}%
\bibitem [{\citenamefont {Ishida}\ \emph {et~al.}(2002)\citenamefont {Ishida},
  \citenamefont {Ozaki}, \citenamefont {Kamatsuka}, \citenamefont {Tou},
  \citenamefont {Kyogaku}, \citenamefont {Kitaoka}, \citenamefont {Tateiwa},
  \citenamefont {Sato}, \citenamefont {Aso}, \citenamefont {Geibel},\ and\
  \citenamefont {Steglich}}]{Ishida2002}%
  \BibitemOpen
  \bibfield  {author} {\bibinfo {author} {\bibfnamefont {K.}~\bibnamefont
  {Ishida}}, \bibinfo {author} {\bibfnamefont {D.}~\bibnamefont {Ozaki}},
  \bibinfo {author} {\bibfnamefont {T.}~\bibnamefont {Kamatsuka}}, \bibinfo
  {author} {\bibfnamefont {H.}~\bibnamefont {Tou}}, \bibinfo {author}
  {\bibfnamefont {M.}~\bibnamefont {Kyogaku}}, \bibinfo {author} {\bibfnamefont
  {Y.}~\bibnamefont {Kitaoka}}, \bibinfo {author} {\bibfnamefont
  {N.}~\bibnamefont {Tateiwa}}, \bibinfo {author} {\bibfnamefont {N.~K.}\
  \bibnamefont {Sato}}, \bibinfo {author} {\bibfnamefont {N.}~\bibnamefont
  {Aso}}, \bibinfo {author} {\bibfnamefont {C.}~\bibnamefont {Geibel}},\ and\
  \bibinfo {author} {\bibfnamefont {F.}~\bibnamefont {Steglich}},\ }\bibfield
  {title} {\bibinfo {title} {{Spin-Triplet Superconductivity in
  $\rm{UNi_{2}Al_{3}}$ Revealed by the ${}^{27}$Al Knight Shift Measurement}},\
  }\href {https://doi.org/10.1103/PhysRevLett.89.037002} {\bibfield  {journal}
  {\bibinfo  {journal} {Phys. Rev. Lett.}\ }\textbf {\bibinfo {volume} {89}},\
  \bibinfo {pages} {037002} (\bibinfo {year} {2002})}\BibitemShut {NoStop}%
\bibitem [{\citenamefont {Mackenzie}\ \emph {et~al.}(2017)\citenamefont
  {Mackenzie}, \citenamefont {Scaffidi}, \citenamefont {Hicks},\ and\
  \citenamefont {Maeno}}]{Mackenzie2017}%
  \BibitemOpen
  \bibfield  {author} {\bibinfo {author} {\bibfnamefont {A.~P.}\ \bibnamefont
  {Mackenzie}}, \bibinfo {author} {\bibfnamefont {T.}~\bibnamefont {Scaffidi}},
  \bibinfo {author} {\bibfnamefont {C.~W.}\ \bibnamefont {Hicks}},\ and\
  \bibinfo {author} {\bibfnamefont {Y.}~\bibnamefont {Maeno}},\ }\bibfield
  {title} {\bibinfo {title} {{Even odder after twenty-three years: the
  superconducting order parameter puzzle of $\rm{Sr_2 Ru O_4}$}},\ }\href
  {https://doi.org/10.1038/s41535-017-0045-4} {\bibfield  {journal} {\bibinfo
  {journal} {npj Quantum Materials}\ }\textbf {\bibinfo {volume} {2}},\
  \bibinfo {pages} {40} (\bibinfo {year} {2017})}\BibitemShut {NoStop}%
\bibitem [{\citenamefont {Pustogow}\ \emph {et~al.}(2019)\citenamefont
  {Pustogow}, \citenamefont {Luo}, \citenamefont {Chronister}, \citenamefont
  {Su}, \citenamefont {Sokolov}, \citenamefont {Jerzembeck}, \citenamefont
  {Mackenzie}, \citenamefont {Hicks}, \citenamefont {Kikugawa}, \citenamefont
  {Raghu}, \citenamefont {Bauer},\ and\ \citenamefont {Brown}}]{Pustogow2019}%
  \BibitemOpen
  \bibfield  {author} {\bibinfo {author} {\bibfnamefont {A.}~\bibnamefont
  {Pustogow}}, \bibinfo {author} {\bibfnamefont {Y.}~\bibnamefont {Luo}},
  \bibinfo {author} {\bibfnamefont {A.}~\bibnamefont {Chronister}}, \bibinfo
  {author} {\bibfnamefont {Y.~S.}\ \bibnamefont {Su}}, \bibinfo {author}
  {\bibfnamefont {D.~A.}\ \bibnamefont {Sokolov}}, \bibinfo {author}
  {\bibfnamefont {F.}~\bibnamefont {Jerzembeck}}, \bibinfo {author}
  {\bibfnamefont {A.~P.}\ \bibnamefont {Mackenzie}}, \bibinfo {author}
  {\bibfnamefont {C.~W.}\ \bibnamefont {Hicks}}, \bibinfo {author}
  {\bibfnamefont {N.}~\bibnamefont {Kikugawa}}, \bibinfo {author}
  {\bibfnamefont {S.}~\bibnamefont {Raghu}}, \bibinfo {author} {\bibfnamefont
  {E.~D.}\ \bibnamefont {Bauer}},\ and\ \bibinfo {author} {\bibfnamefont
  {S.~E.}\ \bibnamefont {Brown}},\ }\bibfield  {title} {\bibinfo {title}
  {{Constraints on the superconducting order parameter in $\rm{Sr_2 Ru O_4}$
  from oxygen-17 nuclear magnetic resonance}},\ }\href
  {https://doi.org/10.1038/s41586-019-1596-2} {\bibfield  {journal} {\bibinfo
  {journal} {Nature}\ }\textbf {\bibinfo {volume} {574}},\ \bibinfo {pages}
  {72} (\bibinfo {year} {2019})}\BibitemShut {NoStop}%
\bibitem [{\citenamefont {Ishida}\ \emph {et~al.}(2020)\citenamefont {Ishida},
  \citenamefont {Manago}, \citenamefont {Kinjo},\ and\ \citenamefont
  {Maeno}}]{Ishida2020}%
  \BibitemOpen
  \bibfield  {author} {\bibinfo {author} {\bibfnamefont {K.}~\bibnamefont
  {Ishida}}, \bibinfo {author} {\bibfnamefont {M.}~\bibnamefont {Manago}},
  \bibinfo {author} {\bibfnamefont {K.}~\bibnamefont {Kinjo}},\ and\ \bibinfo
  {author} {\bibfnamefont {Y.}~\bibnamefont {Maeno}},\ }\bibfield  {title}
  {\bibinfo {title} {{Reduction of the ${}^{17}$O Knight shift in the
  superconducting state and the heat-up effect by NMR pulses on $\rm{Sr_2 Ru
  O_4}$}},\ }\href {https://doi.org/10.7566/JPSJ.89.034712} {\bibfield
  {journal} {\bibinfo  {journal} {J. Phys. Soc. Japan}\ }\textbf {\bibinfo
  {volume} {89}},\ \bibinfo {pages} {1} (\bibinfo {year} {2020})}\BibitemShut
  {NoStop}%
\bibitem [{\citenamefont {Petsch}\ \emph {et~al.}(2020)\citenamefont {Petsch},
  \citenamefont {Zhu}, \citenamefont {Enderle}, \citenamefont {Mao},
  \citenamefont {Maeno}, \citenamefont {Mazin},\ and\ \citenamefont
  {Hayden}}]{Petsch2020}%
  \BibitemOpen
  \bibfield  {author} {\bibinfo {author} {\bibfnamefont {A.~N.}\ \bibnamefont
  {Petsch}}, \bibinfo {author} {\bibfnamefont {M.}~\bibnamefont {Zhu}},
  \bibinfo {author} {\bibfnamefont {M.}~\bibnamefont {Enderle}}, \bibinfo
  {author} {\bibfnamefont {Z.~Q.}\ \bibnamefont {Mao}}, \bibinfo {author}
  {\bibfnamefont {Y.}~\bibnamefont {Maeno}}, \bibinfo {author} {\bibfnamefont
  {I.~I.}\ \bibnamefont {Mazin}},\ and\ \bibinfo {author} {\bibfnamefont
  {S.~M.}\ \bibnamefont {Hayden}},\ }\bibfield  {title} {\bibinfo {title}
  {{Reduction of the Spin Susceptibility in the Superconducting State of
  $\rm{Sr_2 Ru O_4}$ Observed by Polarized Neutron Scattering}},\ }\href
  {https://doi.org/10.1103/PhysRevLett.125.217004} {\bibfield  {journal}
  {\bibinfo  {journal} {Phys. Rev. Lett.}\ }\textbf {\bibinfo {volume} {125}},\
  \bibinfo {pages} {217004} (\bibinfo {year} {2020})}\BibitemShut {NoStop}%
\bibitem [{\citenamefont {Suh}\ \emph {et~al.}(2020)\citenamefont {Suh},
  \citenamefont {Menke}, \citenamefont {Brydon}, \citenamefont {Timm},
  \citenamefont {Ramires},\ and\ \citenamefont {Agterberg}}]{Suh2020}%
  \BibitemOpen
  \bibfield  {author} {\bibinfo {author} {\bibfnamefont {H.~G.}\ \bibnamefont
  {Suh}}, \bibinfo {author} {\bibfnamefont {H.}~\bibnamefont {Menke}}, \bibinfo
  {author} {\bibfnamefont {P.~M.~R.}\ \bibnamefont {Brydon}}, \bibinfo {author}
  {\bibfnamefont {C.}~\bibnamefont {Timm}}, \bibinfo {author} {\bibfnamefont
  {A.}~\bibnamefont {Ramires}},\ and\ \bibinfo {author} {\bibfnamefont {D.~F.}\
  \bibnamefont {Agterberg}},\ }\bibfield  {title} {\bibinfo {title}
  {{Stabilizing even-parity chiral superconductivity in $\rm{Sr_2 Ru O_4}$}},\
  }\href {https://doi.org/10.1103/PhysRevResearch.2.032023} {\bibfield
  {journal} {\bibinfo  {journal} {Phys. Rev. Res.}\ }\textbf {\bibinfo {volume}
  {2}},\ \bibinfo {pages} {032023} (\bibinfo {year} {2020})}\BibitemShut
  {NoStop}%
\bibitem [{\citenamefont {K{\"{a}}ser}\ \emph {et~al.}(2022)\citenamefont
  {K{\"{a}}ser}, \citenamefont {Strand}, \citenamefont {Wentzell},
  \citenamefont {Georges}, \citenamefont {Parcollet},\ and\ \citenamefont
  {Hansmann}}]{Kaeser2022}%
  \BibitemOpen
  \bibfield  {author} {\bibinfo {author} {\bibfnamefont {S.}~\bibnamefont
  {K{\"{a}}ser}}, \bibinfo {author} {\bibfnamefont {H.~U.~R.}\ \bibnamefont
  {Strand}}, \bibinfo {author} {\bibfnamefont {N.}~\bibnamefont {Wentzell}},
  \bibinfo {author} {\bibfnamefont {A.}~\bibnamefont {Georges}}, \bibinfo
  {author} {\bibfnamefont {O.}~\bibnamefont {Parcollet}},\ and\ \bibinfo
  {author} {\bibfnamefont {P.}~\bibnamefont {Hansmann}},\ }\bibfield  {title}
  {\bibinfo {title} {{Interorbital singlet pairing in $\rm{Sr_2 Ru O_4}$: A
  Hund's superconductor}},\ }\href
  {https://doi.org/10.1103/PhysRevB.105.155101} {\bibfield  {journal} {\bibinfo
   {journal} {Phys. Rev. B}\ }\textbf {\bibinfo {volume} {105}},\ \bibinfo
  {pages} {155101} (\bibinfo {year} {2022})}\BibitemShut {NoStop}%
\bibitem [{\citenamefont {Kozii}\ and\ \citenamefont {Fu}(2015)}]{Kozii2015}%
  \BibitemOpen
  \bibfield  {author} {\bibinfo {author} {\bibfnamefont {V.}~\bibnamefont
  {Kozii}}\ and\ \bibinfo {author} {\bibfnamefont {L.}~\bibnamefont {Fu}},\
  }\bibfield  {title} {\bibinfo {title} {{Odd-Parity Superconductivity in the
  Vicinity of Inversion Symmetry Breaking in Spin-Orbit-Coupled Systems}},\
  }\href {https://doi.org/10.1103/PhysRevLett.115.207002} {\bibfield  {journal}
  {\bibinfo  {journal} {Phys. Rev. Lett.}\ }\textbf {\bibinfo {volume} {115}},\
  \bibinfo {pages} {207002} (\bibinfo {year} {2015})}\BibitemShut {NoStop}%
\bibitem [{\citenamefont {Schumann}\ \emph {et~al.}(2020)\citenamefont
  {Schumann}, \citenamefont {Galletti}, \citenamefont {Jeong}, \citenamefont
  {Ahadi}, \citenamefont {Strickland}, \citenamefont {Salmani-Rezaie},\ and\
  \citenamefont {Stemmer}}]{Schumann2020}%
  \BibitemOpen
  \bibfield  {author} {\bibinfo {author} {\bibfnamefont {T.}~\bibnamefont
  {Schumann}}, \bibinfo {author} {\bibfnamefont {L.}~\bibnamefont {Galletti}},
  \bibinfo {author} {\bibfnamefont {H.}~\bibnamefont {Jeong}}, \bibinfo
  {author} {\bibfnamefont {K.}~\bibnamefont {Ahadi}}, \bibinfo {author}
  {\bibfnamefont {W.~M.}\ \bibnamefont {Strickland}}, \bibinfo {author}
  {\bibfnamefont {S.}~\bibnamefont {Salmani-Rezaie}},\ and\ \bibinfo {author}
  {\bibfnamefont {S.}~\bibnamefont {Stemmer}},\ }\bibfield  {title} {\bibinfo
  {title} {{Possible signatures of mixed-parity superconductivity in doped
  polar $\rm{Sr Ti O_3}$ films}},\ }\href
  {https://doi.org/10.1103/PhysRevB.101.100503} {\bibfield  {journal} {\bibinfo
   {journal} {Phys. Rev. B}\ }\textbf {\bibinfo {volume} {101}},\ \bibinfo
  {pages} {100503} (\bibinfo {year} {2020})}\BibitemShut {NoStop}%
\bibitem [{\citenamefont {Wang}\ \emph {et~al.}(2016)\citenamefont {Wang},
  \citenamefont {Cho}, \citenamefont {Hughes},\ and\ \citenamefont
  {Fradkin}}]{Wang2016}%
  \BibitemOpen
  \bibfield  {author} {\bibinfo {author} {\bibfnamefont {Y.}~\bibnamefont
  {Wang}}, \bibinfo {author} {\bibfnamefont {G.~Y.}\ \bibnamefont {Cho}},
  \bibinfo {author} {\bibfnamefont {T.~L.}\ \bibnamefont {Hughes}},\ and\
  \bibinfo {author} {\bibfnamefont {E.}~\bibnamefont {Fradkin}},\ }\bibfield
  {title} {\bibinfo {title} {{Topological superconducting phases from inversion
  symmetry breaking order in spin-orbit-coupled systems}},\ }\href
  {https://doi.org/10.1103/PhysRevB.93.134512} {\bibfield  {journal} {\bibinfo
  {journal} {Phys. Rev. B}\ }\textbf {\bibinfo {volume} {93}},\ \bibinfo
  {pages} {1} (\bibinfo {year} {2016})}\BibitemShut {NoStop}%
\bibitem [{\citenamefont {Wang}\ and\ \citenamefont {Fu}(2017)}]{Wang2017}%
  \BibitemOpen
  \bibfield  {author} {\bibinfo {author} {\bibfnamefont {Y.}~\bibnamefont
  {Wang}}\ and\ \bibinfo {author} {\bibfnamefont {L.}~\bibnamefont {Fu}},\
  }\bibfield  {title} {\bibinfo {title} {{Topological Phase Transitions in
  Multicomponent Superconductors}},\ }\href
  {https://doi.org/10.1103/PhysRevLett.119.187003} {\bibfield  {journal}
  {\bibinfo  {journal} {Phys. Rev. Lett.}\ }\textbf {\bibinfo {volume} {119}},\
  \bibinfo {pages} {187003} (\bibinfo {year} {2017})}\BibitemShut {NoStop}%
\bibitem [{\citenamefont {Khim}\ \emph {et~al.}(2021)\citenamefont {Khim},
  \citenamefont {Landaeta}, \citenamefont {Banda}, \citenamefont {Bannor},
  \citenamefont {Brando}, \citenamefont {Brydon}, \citenamefont {Hafner},
  \citenamefont {K{\"{u}}chler}, \citenamefont {Cardoso-Gil}, \citenamefont
  {Stockert}, \citenamefont {Mackenzie}, \citenamefont {Agterberg},
  \citenamefont {Geibel},\ and\ \citenamefont {Hassinger}}]{Khim2021}%
  \BibitemOpen
  \bibfield  {author} {\bibinfo {author} {\bibfnamefont {S.}~\bibnamefont
  {Khim}}, \bibinfo {author} {\bibfnamefont {J.~F.}\ \bibnamefont {Landaeta}},
  \bibinfo {author} {\bibfnamefont {J.}~\bibnamefont {Banda}}, \bibinfo
  {author} {\bibfnamefont {N.}~\bibnamefont {Bannor}}, \bibinfo {author}
  {\bibfnamefont {M.}~\bibnamefont {Brando}}, \bibinfo {author} {\bibfnamefont
  {P.~M.~R.}\ \bibnamefont {Brydon}}, \bibinfo {author} {\bibfnamefont
  {D.}~\bibnamefont {Hafner}}, \bibinfo {author} {\bibfnamefont
  {R.}~\bibnamefont {K{\"{u}}chler}}, \bibinfo {author} {\bibfnamefont
  {R.}~\bibnamefont {Cardoso-Gil}}, \bibinfo {author} {\bibfnamefont
  {U.}~\bibnamefont {Stockert}}, \bibinfo {author} {\bibfnamefont {A.~P.}\
  \bibnamefont {Mackenzie}}, \bibinfo {author} {\bibfnamefont {D.~F.}\
  \bibnamefont {Agterberg}}, \bibinfo {author} {\bibfnamefont {C.}~\bibnamefont
  {Geibel}},\ and\ \bibinfo {author} {\bibfnamefont {E.}~\bibnamefont
  {Hassinger}},\ }\bibfield  {title} {\bibinfo {title} {{Field-induced
  transition within the superconducting state of $\rm{CeRh_{2}As_{2}}$}},\
  }\href {https://doi.org/10.1126/science.abe7518} {\bibfield  {journal}
  {\bibinfo  {journal} {Science}\ }\textbf {\bibinfo {volume} {373}},\ \bibinfo
  {pages} {1012} (\bibinfo {year} {2021})}\BibitemShut {NoStop}%
\bibitem [{\citenamefont {Hafner}\ \emph {et~al.}(2022)\citenamefont {Hafner},
  \citenamefont {Khanenko}, \citenamefont {Eljaouhari}, \citenamefont
  {K{\"{u}}chler}, \citenamefont {Banda}, \citenamefont {Bannor}, \citenamefont
  {L{\"{u}}hmann}, \citenamefont {Landaeta}, \citenamefont {Mishra},
  \citenamefont {Sheikin}, \citenamefont {Hassinger}, \citenamefont {Khim},
  \citenamefont {Geibel}, \citenamefont {Zwicknagl},\ and\ \citenamefont
  {Brando}}]{Hafner2022}%
  \BibitemOpen
  \bibfield  {author} {\bibinfo {author} {\bibfnamefont {D.}~\bibnamefont
  {Hafner}}, \bibinfo {author} {\bibfnamefont {P.}~\bibnamefont {Khanenko}},
  \bibinfo {author} {\bibfnamefont {E.-O.}\ \bibnamefont {Eljaouhari}},
  \bibinfo {author} {\bibfnamefont {R.}~\bibnamefont {K{\"{u}}chler}}, \bibinfo
  {author} {\bibfnamefont {J.}~\bibnamefont {Banda}}, \bibinfo {author}
  {\bibfnamefont {N.}~\bibnamefont {Bannor}}, \bibinfo {author} {\bibfnamefont
  {T.}~\bibnamefont {L{\"{u}}hmann}}, \bibinfo {author} {\bibfnamefont {J.~F.}\
  \bibnamefont {Landaeta}}, \bibinfo {author} {\bibfnamefont {S.}~\bibnamefont
  {Mishra}}, \bibinfo {author} {\bibfnamefont {I.}~\bibnamefont {Sheikin}},
  \bibinfo {author} {\bibfnamefont {E.}~\bibnamefont {Hassinger}}, \bibinfo
  {author} {\bibfnamefont {S.}~\bibnamefont {Khim}}, \bibinfo {author}
  {\bibfnamefont {C.}~\bibnamefont {Geibel}}, \bibinfo {author} {\bibfnamefont
  {G.}~\bibnamefont {Zwicknagl}},\ and\ \bibinfo {author} {\bibfnamefont
  {M.}~\bibnamefont {Brando}},\ }\bibfield  {title} {\bibinfo {title}
  {{Possible Quadrupole Density Wave in the Superconducting Kondo Lattice
  $\rm{Ce Rh_2 As_2}$}},\ }\href {https://doi.org/10.1103/PhysRevX.12.011023}
  {\bibfield  {journal} {\bibinfo  {journal} {Phys. Rev. X}\ }\textbf {\bibinfo
  {volume} {12}},\ \bibinfo {pages} {011023} (\bibinfo {year}
  {2022})}\BibitemShut {NoStop}%
\bibitem [{\citenamefont {Maruyama}\ \emph {et~al.}(2012)\citenamefont
  {Maruyama}, \citenamefont {Sigrist},\ and\ \citenamefont
  {Yanase}}]{Maruyama2012}%
  \BibitemOpen
  \bibfield  {author} {\bibinfo {author} {\bibfnamefont {D.}~\bibnamefont
  {Maruyama}}, \bibinfo {author} {\bibfnamefont {M.}~\bibnamefont {Sigrist}},\
  and\ \bibinfo {author} {\bibfnamefont {Y.}~\bibnamefont {Yanase}},\
  }\bibfield  {title} {\bibinfo {title} {{Locally non-centrosymmetric
  superconductivity in multilayer systems}},\ }\href
  {https://doi.org/10.1143/JPSJ.81.034702} {\bibfield  {journal} {\bibinfo
  {journal} {J. Phys. Soc. Japan}\ }\textbf {\bibinfo {volume} {81}},\ \bibinfo
  {pages} {1} (\bibinfo {year} {2012})}\BibitemShut {NoStop}%
\bibitem [{\citenamefont {Sigrist}\ \emph {et~al.}(2014)\citenamefont
  {Sigrist}, \citenamefont {Agterberg}, \citenamefont {Fischer}, \citenamefont
  {Goryo}, \citenamefont {Loder}, \citenamefont {Rhim}, \citenamefont
  {Maruyama}, \citenamefont {Yanase}, \citenamefont {Yoshida},\ and\
  \citenamefont {Youn}}]{Sigrist2014}%
  \BibitemOpen
  \bibfield  {author} {\bibinfo {author} {\bibfnamefont {M.}~\bibnamefont
  {Sigrist}}, \bibinfo {author} {\bibfnamefont {D.~F.}\ \bibnamefont
  {Agterberg}}, \bibinfo {author} {\bibfnamefont {M.~H.}\ \bibnamefont
  {Fischer}}, \bibinfo {author} {\bibfnamefont {J.}~\bibnamefont {Goryo}},
  \bibinfo {author} {\bibfnamefont {F.}~\bibnamefont {Loder}}, \bibinfo
  {author} {\bibfnamefont {S.~H.}\ \bibnamefont {Rhim}}, \bibinfo {author}
  {\bibfnamefont {D.}~\bibnamefont {Maruyama}}, \bibinfo {author}
  {\bibfnamefont {Y.}~\bibnamefont {Yanase}}, \bibinfo {author} {\bibfnamefont
  {T.}~\bibnamefont {Yoshida}},\ and\ \bibinfo {author} {\bibfnamefont {S.~J.}\
  \bibnamefont {Youn}},\ }\bibfield  {title} {\bibinfo {title}
  {{Superconductors with staggered non-centrosymmetricity}},\ }\href
  {https://doi.org/10.7566/JPSJ.83.061014} {\bibfield  {journal} {\bibinfo
  {journal} {J. Phys. Soc. Japan}\ }\textbf {\bibinfo {volume} {83}},\ \bibinfo
  {pages} {1} (\bibinfo {year} {2014})}\BibitemShut {NoStop}%
\bibitem [{\citenamefont {Clogston}(1962)}]{Clogston1962}%
  \BibitemOpen
  \bibfield  {author} {\bibinfo {author} {\bibfnamefont {A.~M.}\ \bibnamefont
  {Clogston}},\ }\bibfield  {title} {\bibinfo {title} {{Upper limit for the
  critical field in hard superconductors}},\ }\href
  {https://doi.org/10.1103/PhysRevLett.9.266} {\bibfield  {journal} {\bibinfo
  {journal} {Phys. Rev. Lett.}\ }\textbf {\bibinfo {volume} {9}},\ \bibinfo
  {pages} {266} (\bibinfo {year} {1962})}\BibitemShut {NoStop}%
\bibitem [{\citenamefont {Sarma}(1963)}]{Sarma1963}%
  \BibitemOpen
  \bibfield  {author} {\bibinfo {author} {\bibfnamefont {G.}~\bibnamefont
  {Sarma}},\ }\bibfield  {title} {\bibinfo {title} {{On the influence of a
  uniform exchange field acting on the spins of the conduction electrons in a
  superconductor}},\ }\href {https://doi.org/10.1016/0022-3697(63)90007-6}
  {\bibfield  {journal} {\bibinfo  {journal} {J. Phys. Chem. Solids}\ }\textbf
  {\bibinfo {volume} {24}},\ \bibinfo {pages} {1029} (\bibinfo {year}
  {1963})}\BibitemShut {NoStop}%
\bibitem [{\citenamefont {Maeno}\ \emph {et~al.}(2012)\citenamefont {Maeno},
  \citenamefont {Kittaka}, \citenamefont {Nomura}, \citenamefont {Yonezawa},\
  and\ \citenamefont {Ishida}}]{Maeno2012}%
  \BibitemOpen
  \bibfield  {author} {\bibinfo {author} {\bibfnamefont {Y.}~\bibnamefont
  {Maeno}}, \bibinfo {author} {\bibfnamefont {S.}~\bibnamefont {Kittaka}},
  \bibinfo {author} {\bibfnamefont {T.}~\bibnamefont {Nomura}}, \bibinfo
  {author} {\bibfnamefont {S.}~\bibnamefont {Yonezawa}},\ and\ \bibinfo
  {author} {\bibfnamefont {K.}~\bibnamefont {Ishida}},\ }\bibfield  {title}
  {\bibinfo {title} {{Evaluation of spin-triplet superconductivity in $\rm{Sr_2
  Ru O_4}$}},\ }\href {https://doi.org/10.1143/JPSJ.81.011009} {\bibfield
  {journal} {\bibinfo  {journal} {J. Phys. Soc. Japan}\ }\textbf {\bibinfo
  {volume} {81}},\ \bibinfo {pages} {1} (\bibinfo {year} {2012})}\BibitemShut
  {NoStop}%
\bibitem [{\citenamefont {Cavanagh}\ \emph {et~al.}(2022)\citenamefont
  {Cavanagh}, \citenamefont {Shishidou}, \citenamefont {Weinert}, \citenamefont
  {Brydon},\ and\ \citenamefont {Agterberg}}]{Cavanagh2022}%
  \BibitemOpen
  \bibfield  {author} {\bibinfo {author} {\bibfnamefont {D.~C.}\ \bibnamefont
  {Cavanagh}}, \bibinfo {author} {\bibfnamefont {T.}~\bibnamefont {Shishidou}},
  \bibinfo {author} {\bibfnamefont {M.}~\bibnamefont {Weinert}}, \bibinfo
  {author} {\bibfnamefont {P.~M.~R.}\ \bibnamefont {Brydon}},\ and\ \bibinfo
  {author} {\bibfnamefont {D.~F.}\ \bibnamefont {Agterberg}},\ }\bibfield
  {title} {\bibinfo {title} {{Nonsymmorphic symmetry and field-driven
  odd-parity pairing in $\rm{CeRh_{2}As_{2}}$}},\ }\href
  {https://doi.org/10.1103/PhysRevB.105.L020505} {\bibfield  {journal}
  {\bibinfo  {journal} {Phys. Rev. B}\ }\textbf {\bibinfo {volume} {105}},\
  \bibinfo {pages} {L020505} (\bibinfo {year} {2022})}\BibitemShut {NoStop}%
\bibitem [{\citenamefont {M{\"{o}}ckli}\ and\ \citenamefont
  {Ramires}(2021)}]{Mockli2021}%
  \BibitemOpen
  \bibfield  {author} {\bibinfo {author} {\bibfnamefont {D.}~\bibnamefont
  {M{\"{o}}ckli}}\ and\ \bibinfo {author} {\bibfnamefont {A.}~\bibnamefont
  {Ramires}},\ }\bibfield  {title} {\bibinfo {title} {{Two scenarios for
  superconductivity in $\rm{CeRh_{2}As_{2}}$}},\ }\href
  {https://doi.org/10.1103/PhysRevResearch.3.023204} {\bibfield  {journal}
  {\bibinfo  {journal} {Phys. Rev. Res.}\ }\textbf {\bibinfo {volume} {3}},\
  \bibinfo {pages} {023204} (\bibinfo {year} {2021})}\BibitemShut {NoStop}%
\bibitem [{\citenamefont {Yoshida}\ \emph {et~al.}(2012)\citenamefont
  {Yoshida}, \citenamefont {Sigrist},\ and\ \citenamefont
  {Yanase}}]{Yoshida2012}%
  \BibitemOpen
  \bibfield  {author} {\bibinfo {author} {\bibfnamefont {T.}~\bibnamefont
  {Yoshida}}, \bibinfo {author} {\bibfnamefont {M.}~\bibnamefont {Sigrist}},\
  and\ \bibinfo {author} {\bibfnamefont {Y.}~\bibnamefont {Yanase}},\
  }\bibfield  {title} {\bibinfo {title} {{Pair-density wave states through
  spin-orbit coupling in multilayer superconductors}},\ }\href
  {https://doi.org/10.1103/PhysRevB.86.134514} {\bibfield  {journal} {\bibinfo
  {journal} {Phys. Rev. B}\ }\textbf {\bibinfo {volume} {86}},\ \bibinfo
  {pages} {1} (\bibinfo {year} {2012})}\BibitemShut {NoStop}%
\bibitem [{\citenamefont {Machida}(2022)}]{Machida2022}%
  \BibitemOpen
  \bibfield  {author} {\bibinfo {author} {\bibfnamefont {K.}~\bibnamefont
  {Machida}},\ }\bibfield  {title} {\bibinfo {title} {{Violation of the
  Pauli-Clogston limit in a heavy Fermion superconductor $\rm{Ce
  Rh_{2}As_{2}}$-Duality of itinerant and localized 4f electrons}},\ }\href
  {https://doi.org/10.1103/PhysRevB.106.184509} {\bibfield  {journal} {\bibinfo
   {journal} {Phys. Rev. B}\ }\textbf {\bibinfo {volume} {184509}},\ \bibinfo
  {pages} {1} (\bibinfo {year} {2022})}\BibitemShut {NoStop}%
\bibitem [{\citenamefont {Sauls}(2022)}]{Sauls2022}%
  \BibitemOpen
  \bibfield  {author} {\bibinfo {author} {\bibfnamefont {J.~A.}\ \bibnamefont
  {Sauls}},\ }\bibfield  {title} {\bibinfo {title} {{On the Excitations of a
  Balian-Werthamer Superconductor}},\ }\href
  {https://doi.org/10.1007/s10909-022-02748-2} {\bibfield  {journal} {\bibinfo
  {journal} {J Low Temp Phys}\ }\textbf {\bibinfo {volume} {208}},\ \bibinfo
  {pages} {87} (\bibinfo {year} {2022})}\BibitemShut {NoStop}%
\bibitem [{\citenamefont {Bardasis}\ and\ \citenamefont
  {Schrieffer}(1961)}]{Bardasis1961}%
  \BibitemOpen
  \bibfield  {author} {\bibinfo {author} {\bibfnamefont {A.}~\bibnamefont
  {Bardasis}}\ and\ \bibinfo {author} {\bibfnamefont {J.~R.}\ \bibnamefont
  {Schrieffer}},\ }\bibfield  {title} {\bibinfo {title} {{Excitons and Plasmons
  in Superconductors}},\ }\href {https://doi.org/10.1103/PhysRev.121.1050}
  {\bibfield  {journal} {\bibinfo  {journal} {Phys. Rev.}\ }\textbf {\bibinfo
  {volume} {121}},\ \bibinfo {pages} {1050} (\bibinfo {year}
  {1961})}\BibitemShut {NoStop}%
\bibitem [{\citenamefont {B{\"{o}}hm}\ \emph {et~al.}(2014)\citenamefont
  {B{\"{o}}hm}, \citenamefont {Kemper}, \citenamefont {Moritz}, \citenamefont
  {Kretzschmar}, \citenamefont {Muschler}, \citenamefont {Eiter}, \citenamefont
  {Hackl}, \citenamefont {Devereaux}, \citenamefont {Scalapino},\ and\
  \citenamefont {Wen}}]{Bohm2014}%
  \BibitemOpen
  \bibfield  {author} {\bibinfo {author} {\bibfnamefont {T.}~\bibnamefont
  {B{\"{o}}hm}}, \bibinfo {author} {\bibfnamefont {A.~F.}\ \bibnamefont
  {Kemper}}, \bibinfo {author} {\bibfnamefont {B.}~\bibnamefont {Moritz}},
  \bibinfo {author} {\bibfnamefont {F.}~\bibnamefont {Kretzschmar}}, \bibinfo
  {author} {\bibfnamefont {B.}~\bibnamefont {Muschler}}, \bibinfo {author}
  {\bibfnamefont {H.~M.}\ \bibnamefont {Eiter}}, \bibinfo {author}
  {\bibfnamefont {R.}~\bibnamefont {Hackl}}, \bibinfo {author} {\bibfnamefont
  {T.~P.}\ \bibnamefont {Devereaux}}, \bibinfo {author} {\bibfnamefont {D.~J.}\
  \bibnamefont {Scalapino}},\ and\ \bibinfo {author} {\bibfnamefont {H.~H.}\
  \bibnamefont {Wen}},\ }\bibfield  {title} {\bibinfo {title} {Balancing act:
  Evidence for a strong subdominant d-wave pairing channel in $\rm{Ba_{0.6}
  K_{0.4} Fe_{2}As_{2}}$},\ }\href {https://doi.org/10.1103/PhysRevX.4.041046}
  {\bibfield  {journal} {\bibinfo  {journal} {Phys. Rev. X}\ }\textbf {\bibinfo
  {volume} {4}},\ \bibinfo {pages} {1} (\bibinfo {year} {2014})}\BibitemShut
  {NoStop}%
\bibitem [{\citenamefont {He}\ \emph {et~al.}(2020)\citenamefont {He},
  \citenamefont {Li}, \citenamefont {Jost}, \citenamefont {Baum}, \citenamefont
  {Shen}, \citenamefont {Dong}, \citenamefont {Zhao},\ and\ \citenamefont
  {Hackl}}]{He2020}%
  \BibitemOpen
  \bibfield  {author} {\bibinfo {author} {\bibfnamefont {G.}~\bibnamefont
  {He}}, \bibinfo {author} {\bibfnamefont {D.}~\bibnamefont {Li}}, \bibinfo
  {author} {\bibfnamefont {D.}~\bibnamefont {Jost}}, \bibinfo {author}
  {\bibfnamefont {A.}~\bibnamefont {Baum}}, \bibinfo {author} {\bibfnamefont
  {P.~P.}\ \bibnamefont {Shen}}, \bibinfo {author} {\bibfnamefont {X.~L.}\
  \bibnamefont {Dong}}, \bibinfo {author} {\bibfnamefont {Z.~X.}\ \bibnamefont
  {Zhao}},\ and\ \bibinfo {author} {\bibfnamefont {R.}~\bibnamefont {Hackl}},\
  }\bibfield  {title} {\bibinfo {title} {{Raman Study of Cooper Pairing
  Instabilities in $\rm{Li_{1-x}Fe_{x}OHFeSe}$}},\ }\href
  {https://doi.org/10.1103/PhysRevLett.125.217002} {\bibfield  {journal}
  {\bibinfo  {journal} {Phys. Rev. Lett.}\ }\textbf {\bibinfo {volume} {125}},\
  \bibinfo {pages} {1} (\bibinfo {year} {2020})}\BibitemShut {NoStop}%
\bibitem [{\citenamefont {Anderson}(1958)}]{Anderson1958}%
  \BibitemOpen
  \bibfield  {author} {\bibinfo {author} {\bibfnamefont {P.~W.}\ \bibnamefont
  {Anderson}},\ }\bibfield  {title} {\bibinfo {title} {{Random-phase
  approximation in the theory of superconductivity}},\ }\href
  {https://doi.org/10.1103/PhysRev.112.1900} {\bibfield  {journal} {\bibinfo
  {journal} {Phys. Rev.}\ }\textbf {\bibinfo {volume} {112}},\ \bibinfo {pages}
  {1900} (\bibinfo {year} {1958})}\BibitemShut {NoStop}%
\bibitem [{\citenamefont {Rickayzen}(1959)}]{Rickayzen1959}%
  \BibitemOpen
  \bibfield  {author} {\bibinfo {author} {\bibfnamefont {G.}~\bibnamefont
  {Rickayzen}},\ }\bibfield  {title} {\bibinfo {title} {{Collective excitations
  in the theory of superconductivity}},\ }\href
  {https://doi.org/10.1103/PhysRev.115.795} {\bibfield  {journal} {\bibinfo
  {journal} {Phys. Rev.}\ }\textbf {\bibinfo {volume} {115}},\ \bibinfo {pages}
  {795} (\bibinfo {year} {1959})}\BibitemShut {NoStop}%
\bibitem [{\citenamefont {M{\"{o}}ckli}\ \emph {et~al.}(2018)\citenamefont
  {M{\"{o}}ckli}, \citenamefont {Yanase},\ and\ \citenamefont
  {Sigrist}}]{Mockli2018}%
  \BibitemOpen
  \bibfield  {author} {\bibinfo {author} {\bibfnamefont {D.}~\bibnamefont
  {M{\"{o}}ckli}}, \bibinfo {author} {\bibfnamefont {Y.}~\bibnamefont
  {Yanase}},\ and\ \bibinfo {author} {\bibfnamefont {M.}~\bibnamefont
  {Sigrist}},\ }\bibfield  {title} {\bibinfo {title} {{Orbitally limited
  pair-density-wave phase of multilayer superconductors}},\ }\href
  {https://doi.org/10.1103/PhysRevB.97.144508} {\bibfield  {journal} {\bibinfo
  {journal} {Phys. Rev. B}\ }\textbf {\bibinfo {volume} {97}},\ \bibinfo
  {pages} {144508} (\bibinfo {year} {2018})}\BibitemShut {NoStop}%
\bibitem [{\citenamefont {Balian}\ and\ \citenamefont
  {Werthamer}(1963)}]{Balian1963}%
  \BibitemOpen
  \bibfield  {author} {\bibinfo {author} {\bibfnamefont {R.}~\bibnamefont
  {Balian}}\ and\ \bibinfo {author} {\bibfnamefont {N.~R.}\ \bibnamefont
  {Werthamer}},\ }\bibfield  {title} {\bibinfo {title} {Superconductivity with
  pairs in a relative \emph{p} wave},\ }\href
  {https://doi.org/10.1103/PhysRev.131.1553} {\bibfield  {journal} {\bibinfo
  {journal} {Phys. Rev.}\ }\textbf {\bibinfo {volume} {131}},\ \bibinfo {pages}
  {1553} (\bibinfo {year} {1963})}\BibitemShut {NoStop}%
\bibitem [{\citenamefont {Nambu}(1960)}]{Nambu1960}%
  \BibitemOpen
  \bibfield  {author} {\bibinfo {author} {\bibfnamefont {Y.}~\bibnamefont
  {Nambu}},\ }\bibfield  {title} {\bibinfo {title} {{Quasi-Particles and Gauge
  Invariance in the Theory of Superconductivity}},\ }\href
  {https://doi.org/10.1103/PhysRev.117.648} {\bibfield  {journal} {\bibinfo
  {journal} {Phys. Rev.}\ }\textbf {\bibinfo {volume} {117}},\ \bibinfo {pages}
  {648} (\bibinfo {year} {1960})}\BibitemShut {NoStop}%
\bibitem [{\citenamefont {Skurativska}\ \emph {et~al.}(2021)\citenamefont
  {Skurativska}, \citenamefont {Sigrist},\ and\ \citenamefont
  {Fischer}}]{Skurativska2021}%
  \BibitemOpen
  \bibfield  {author} {\bibinfo {author} {\bibfnamefont {A.}~\bibnamefont
  {Skurativska}}, \bibinfo {author} {\bibfnamefont {M.}~\bibnamefont
  {Sigrist}},\ and\ \bibinfo {author} {\bibfnamefont {M.~H.}\ \bibnamefont
  {Fischer}},\ }\bibfield  {title} {\bibinfo {title} {{Spin response and
  topology of a staggered-Rashba superconductor}},\ }\href
  {https://doi.org/10.1103/PhysRevResearch.3.033133} {\bibfield  {journal}
  {\bibinfo  {journal} {Phys. Rev. Res.}\ }\textbf {\bibinfo {volume} {3}},\
  \bibinfo {pages} {033133} (\bibinfo {year} {2021})}\BibitemShut {NoStop}%
\bibitem [{\citenamefont {Landaeta}\ \emph {et~al.}(2022)\citenamefont
  {Landaeta}, \citenamefont {Khanenko}, \citenamefont {Cavanagh}, \citenamefont
  {Geibel}, \citenamefont {Khim}, \citenamefont {Mishra}, \citenamefont
  {Sheikin}, \citenamefont {Brydon}, \citenamefont {Agterberg}, \citenamefont
  {Brando},\ and\ \citenamefont {Hassinger}}]{Landaeta2022}%
  \BibitemOpen
  \bibfield  {author} {\bibinfo {author} {\bibfnamefont {J.~F.}\ \bibnamefont
  {Landaeta}}, \bibinfo {author} {\bibfnamefont {P.}~\bibnamefont {Khanenko}},
  \bibinfo {author} {\bibfnamefont {D.~C.}\ \bibnamefont {Cavanagh}}, \bibinfo
  {author} {\bibfnamefont {C.}~\bibnamefont {Geibel}}, \bibinfo {author}
  {\bibfnamefont {S.}~\bibnamefont {Khim}}, \bibinfo {author} {\bibfnamefont
  {S.}~\bibnamefont {Mishra}}, \bibinfo {author} {\bibfnamefont
  {I.}~\bibnamefont {Sheikin}}, \bibinfo {author} {\bibfnamefont {P.~M.~R.}\
  \bibnamefont {Brydon}}, \bibinfo {author} {\bibfnamefont {D.~F.}\
  \bibnamefont {Agterberg}}, \bibinfo {author} {\bibfnamefont {M.}~\bibnamefont
  {Brando}},\ and\ \bibinfo {author} {\bibfnamefont {E.}~\bibnamefont
  {Hassinger}},\ }\bibfield  {title} {\bibinfo {title} {{Field-Angle Dependence
  Reveals Odd-Parity Superconductivity in $\rm{CeRh_{2}As_{2}}$}},\ }\href
  {https://doi.org/10.1103/PhysRevX.12.031001} {\bibfield  {journal} {\bibinfo
  {journal} {Phys. Rev. X}\ }\textbf {\bibinfo {volume} {12}},\ \bibinfo
  {pages} {031001} (\bibinfo {year} {2022})}\BibitemShut {NoStop}%
\bibitem [{\citenamefont {Landau}\ and\ \citenamefont
  {Lifshitz}(2013)}]{LandauBook}%
  \BibitemOpen
  \bibfield  {author} {\bibinfo {author} {\bibfnamefont {L.~D.}\ \bibnamefont
  {Landau}}\ and\ \bibinfo {author} {\bibfnamefont {E.~M.}\ \bibnamefont
  {Lifshitz}},\ }\href@noop {} {\emph {\bibinfo {title} {Statistical Physics:
  Volume 5}}},\ Vol.~\bibinfo {volume} {5}\ (\bibinfo  {publisher} {Elsevier},\
  \bibinfo {year} {2013})\BibitemShut {NoStop}%
\bibitem [{\citenamefont {Boyack}\ and\ \citenamefont
  {Lopes}(2020)}]{Boyack2020}%
  \BibitemOpen
  \bibfield  {author} {\bibinfo {author} {\bibfnamefont {R.}~\bibnamefont
  {Boyack}}\ and\ \bibinfo {author} {\bibfnamefont {P.~L.}\ \bibnamefont
  {Lopes}},\ }\bibfield  {title} {\bibinfo {title} {{Electromagnetic response
  of superconductors in the presence of multiple collective modes}},\ }\href
  {https://doi.org/10.1103/PhysRevB.101.094509} {\bibfield  {journal} {\bibinfo
   {journal} {Phys. Rev. B}\ }\textbf {\bibinfo {volume} {101}},\ \bibinfo
  {pages} {94509} (\bibinfo {year} {2020})}\BibitemShut {NoStop}%
\bibitem [{\citenamefont {Maiti}\ and\ \citenamefont
  {Hirschfeld}(2015)}]{Maiti2015}%
  \BibitemOpen
  \bibfield  {author} {\bibinfo {author} {\bibfnamefont {S.}~\bibnamefont
  {Maiti}}\ and\ \bibinfo {author} {\bibfnamefont {P.~J.}\ \bibnamefont
  {Hirschfeld}},\ }\bibfield  {title} {\bibinfo {title} {{Collective modes in
  superconductors with competing s- and d-wave interactions}},\ }\href
  {https://doi.org/10.1103/PhysRevB.92.094506} {\bibfield  {journal} {\bibinfo
  {journal} {Phys. Rev. B}\ }\textbf {\bibinfo {volume} {92}},\ \bibinfo
  {pages} {094506} (\bibinfo {year} {2015})}\BibitemShut {NoStop}%
\bibitem [{\citenamefont {Giuliani}\ and\ \citenamefont
  {Vignale}(2005)}]{giuliani_vignale_2005}%
  \BibitemOpen
  \bibfield  {author} {\bibinfo {author} {\bibfnamefont {G.}~\bibnamefont
  {Giuliani}}\ and\ \bibinfo {author} {\bibfnamefont {G.}~\bibnamefont
  {Vignale}},\ }\href {https://doi.org/10.1017/CBO9780511619915} {\emph
  {\bibinfo {title} {Quantum Theory of the Electron Liquid}}}\ (\bibinfo
  {publisher} {Cambridge University Press},\ \bibinfo {year}
  {2005})\BibitemShut {NoStop}%
\bibitem [{\citenamefont {Ahn}\ and\ \citenamefont {Nagaosa}(2021)}]{Ahn2021}%
  \BibitemOpen
  \bibfield  {author} {\bibinfo {author} {\bibfnamefont {J.}~\bibnamefont
  {Ahn}}\ and\ \bibinfo {author} {\bibfnamefont {N.}~\bibnamefont {Nagaosa}},\
  }\bibfield  {title} {\bibinfo {title} {{Theory of optical responses in clean
  multi-band superconductors}},\ }\href
  {https://doi.org/10.1038/s41467-021-21905-x} {\bibfield  {journal} {\bibinfo
  {journal} {Nat. Commun.}\ }\textbf {\bibinfo {volume} {12}},\ \bibinfo
  {pages} {1617} (\bibinfo {year} {2021})}\BibitemShut {NoStop}%
\bibitem [{\citenamefont {Kamatani}\ \emph {et~al.}(2022)\citenamefont
  {Kamatani}, \citenamefont {Kitamura}, \citenamefont {Tsuji}, \citenamefont
  {Shimano},\ and\ \citenamefont {Morimoto}}]{Kamatani2022}%
  \BibitemOpen
  \bibfield  {author} {\bibinfo {author} {\bibfnamefont {T.}~\bibnamefont
  {Kamatani}}, \bibinfo {author} {\bibfnamefont {S.}~\bibnamefont {Kitamura}},
  \bibinfo {author} {\bibfnamefont {N.}~\bibnamefont {Tsuji}}, \bibinfo
  {author} {\bibfnamefont {R.}~\bibnamefont {Shimano}},\ and\ \bibinfo {author}
  {\bibfnamefont {T.}~\bibnamefont {Morimoto}},\ }\bibfield  {title} {\bibinfo
  {title} {{Optical response of the Leggett mode in multiband superconductors
  in the linear response regime}},\ }\href
  {https://doi.org/10.1103/PhysRevB.105.094520} {\bibfield  {journal} {\bibinfo
   {journal} {Phys. Rev. B}\ }\textbf {\bibinfo {volume} {105}},\ \bibinfo
  {pages} {094520} (\bibinfo {year} {2022})}\BibitemShut {NoStop}%
\bibitem [{\citenamefont {Venderbos}\ \emph {et~al.}(2016)\citenamefont
  {Venderbos}, \citenamefont {Kozii},\ and\ \citenamefont
  {Fu}}]{Venderbos2016}%
  \BibitemOpen
  \bibfield  {author} {\bibinfo {author} {\bibfnamefont {J.~W.~F.}\
  \bibnamefont {Venderbos}}, \bibinfo {author} {\bibfnamefont {V.}~\bibnamefont
  {Kozii}},\ and\ \bibinfo {author} {\bibfnamefont {L.}~\bibnamefont {Fu}},\
  }\bibfield  {title} {\bibinfo {title} {{Odd-parity superconductors with
  two-component order parameters: Nematic and chiral, full gap, and Majorana
  node}},\ }\href {https://doi.org/10.1103/PhysRevB.94.180504} {\bibfield
  {journal} {\bibinfo  {journal} {Phys. Rev. B}\ }\textbf {\bibinfo {volume}
  {94}},\ \bibinfo {pages} {180504} (\bibinfo {year} {2016})}\BibitemShut
  {NoStop}%
\bibitem [{\citenamefont {Kibune}\ \emph {et~al.}(2022)\citenamefont {Kibune},
  \citenamefont {Kitagawa}, \citenamefont {Kinjo}, \citenamefont {Ogata},
  \citenamefont {Manago}, \citenamefont {Taniguchi}, \citenamefont {Ishida},
  \citenamefont {Brando}, \citenamefont {Hassinger}, \citenamefont {Rosner},
  \citenamefont {Geibel},\ and\ \citenamefont {Khim}}]{Kibune2022}%
  \BibitemOpen
  \bibfield  {author} {\bibinfo {author} {\bibfnamefont {M.}~\bibnamefont
  {Kibune}}, \bibinfo {author} {\bibfnamefont {S.}~\bibnamefont {Kitagawa}},
  \bibinfo {author} {\bibfnamefont {K.}~\bibnamefont {Kinjo}}, \bibinfo
  {author} {\bibfnamefont {S.}~\bibnamefont {Ogata}}, \bibinfo {author}
  {\bibfnamefont {M.}~\bibnamefont {Manago}}, \bibinfo {author} {\bibfnamefont
  {T.}~\bibnamefont {Taniguchi}}, \bibinfo {author} {\bibfnamefont
  {K.}~\bibnamefont {Ishida}}, \bibinfo {author} {\bibfnamefont
  {M.}~\bibnamefont {Brando}}, \bibinfo {author} {\bibfnamefont
  {E.}~\bibnamefont {Hassinger}}, \bibinfo {author} {\bibfnamefont
  {H.}~\bibnamefont {Rosner}}, \bibinfo {author} {\bibfnamefont
  {C.}~\bibnamefont {Geibel}},\ and\ \bibinfo {author} {\bibfnamefont
  {S.}~\bibnamefont {Khim}},\ }\bibfield  {title} {\bibinfo {title}
  {{Observation of Antiferromagnetic Order as Odd-Parity Multipoles inside the
  Superconducting Phase in $\rm{Ce Rh_{2}As_{2}}$}},\ }\href
  {https://doi.org/10.1103/PhysRevLett.128.057002} {\bibfield  {journal}
  {\bibinfo  {journal} {Phys. Rev. Lett.}\ }\textbf {\bibinfo {volume} {128}},\
  \bibinfo {pages} {057002} (\bibinfo {year} {2022})}\BibitemShut {NoStop}%
\bibitem [{\citenamefont {Hackner}\ and\ \citenamefont
  {Brydon}()}]{Hackner2022}%
  \BibitemOpen
  \bibfield  {author} {\bibinfo {author} {\bibfnamefont {N.~A.}\ \bibnamefont
  {Hackner}}\ and\ \bibinfo {author} {\bibfnamefont {P.~M.~R.}\ \bibnamefont
  {Brydon}},\ }\bibinfo {note} {(unpublished)}\BibitemShut {NoStop}%
\bibitem [{\citenamefont {Tewari}\ \emph {et~al.}(2011)\citenamefont {Tewari},
  \citenamefont {Stanescu}, \citenamefont {Sau},\ and\ \citenamefont {{Das
  Sarma}}}]{Tewari2011}%
  \BibitemOpen
  \bibfield  {author} {\bibinfo {author} {\bibfnamefont {S.}~\bibnamefont
  {Tewari}}, \bibinfo {author} {\bibfnamefont {T.~D.}\ \bibnamefont
  {Stanescu}}, \bibinfo {author} {\bibfnamefont {J.~D.}\ \bibnamefont {Sau}},\
  and\ \bibinfo {author} {\bibfnamefont {S.}~\bibnamefont {{Das Sarma}}},\
  }\bibfield  {title} {\bibinfo {title} {{Topologically non-trivial
  superconductivity in spin-orbit-coupled systems: bulk phases and quantum
  phase transitions}},\ }\href {https://doi.org/10.1088/1367-2630/13/6/065004}
  {\bibfield  {journal} {\bibinfo  {journal} {New J. Phys.}\ }\textbf {\bibinfo
  {volume} {13}},\ \bibinfo {pages} {065004} (\bibinfo {year}
  {2011})}\BibitemShut {NoStop}%
\bibitem [{\citenamefont {Sauls}\ and\ \citenamefont
  {Mizushima}(2017)}]{Sauls2017}%
  \BibitemOpen
  \bibfield  {author} {\bibinfo {author} {\bibfnamefont {J.~A.}\ \bibnamefont
  {Sauls}}\ and\ \bibinfo {author} {\bibfnamefont {T.}~\bibnamefont
  {Mizushima}},\ }\bibfield  {title} {\bibinfo {title} {{On the Nambu
  fermion-boson relations for superfluid $\rm{{}^{3}He}$}},\ }\href
  {https://doi.org/10.1103/PhysRevB.95.094515} {\bibfield  {journal} {\bibinfo
  {journal} {Phys. Rev. B}\ }\textbf {\bibinfo {volume} {95}},\ \bibinfo
  {pages} {094515} (\bibinfo {year} {2017})}\BibitemShut {NoStop}%
\bibitem [{\citenamefont {Nambu}(1985)}]{Nambu1985}%
  \BibitemOpen
  \bibfield  {author} {\bibinfo {author} {\bibfnamefont {Y.}~\bibnamefont
  {Nambu}},\ }\bibfield  {title} {\bibinfo {title} {{Fermion-Boson relations in
  BCS-type theories}},\ }\href
  {https://doi.org/https://doi.org/10.1016/0167-2789(85)90157-5} {\bibfield
  {journal} {\bibinfo  {journal} {Physica D: Nonlinear Phenomena}\ }\textbf
  {\bibinfo {volume} {15}},\ \bibinfo {pages} {147} (\bibinfo {year}
  {1985})}\BibitemShut {NoStop}%
\end{thebibliography}%
\end{document}